\documentclass[a4paper,11pt]{article}
\usepackage[utf8]{inputenc}
\usepackage{mathtools}
\usepackage{amssymb,amsmath}
\usepackage[affil-it]{authblk}
\usepackage{float}
\usepackage{subcaption}
\usepackage{verbatim}
\usepackage{enumitem}
\usepackage{authblk}
\usepackage{longtable}
\usepackage{import}
\usepackage{geometry} 
\usepackage{breqn}

\usepackage{titlesec}
\usepackage[titletoc,toc,title]{appendix}
\titlespacing*{\section}{0pt}{25pt}{17pt}
\titlespacing*{\subsection}{0pt}{15pt}{12pt}

\usepackage{array} 
\newcolumntype{H}{>{\setbox0=\hbox\bgroup}c<{\egroup}@{}} 

\usepackage{multirow}
\usepackage{comment}

\DeclareMathAlphabet{\mathbit}{OT1}{cmr}{bx}{it}

\usepackage{pdflscape}
\usepackage{tabu}
\usepackage{adjustbox}

\usepackage{graphicx}
\graphicspath{{figures/}}

\usepackage[bottom, multiple]{footmisc} 
\usepackage{natbib}
\usepackage{xr-hyper} 
\usepackage{hyperref}
\hypersetup{colorlinks=true,urlcolor=blue,citecolor=blue,linkcolor=blue}    
\makeatletter
\g@addto@macro\normalsize{%
	\setlength\abovedisplayskip{10pt}
	\setlength\belowdisplayskip{10pt}
	\setlength\abovedisplayshortskip{10pt}
	\setlength\belowdisplayshortskip{10pt}
}
\makeatother

\title{Financial-Cycle Ratios and Medium-Term Predictions of GDP: Evidence from the United States} 
\author{Graziano Moramarco\thanks{University of Bologna, Department of Economics - Piazza Scaravilli 2, 40126 Bologna, Italy. \newline E-mail: graziano.moramarco@unibo.it. 
I gratefully acknowledge financial support from the Italian Ministry of Education, Universities and Research (MIUR), PRIN 2017 Grant 2017TA7TYC.}}
\date{}

\begin{document}

	\maketitle
	
	\begin{abstract}  
		Using a large quarterly macroeconomic dataset for the period 1960-2017, we document the ability of specific financial ratios from the housing market and firms' aggregate balance sheets to predict GDP over medium-term horizons in the United States. 
		A cyclically-adjusted house price-to-rent ratio and the liabilities-to-income ratio of the nonfinancial noncorporate business sector provide the best in-sample and out-of-sample predictions of GDP growth over horizons of 1-5 years, based on a wide variety of rankings.
		Small forecasting models that include these indicators outperform popular high-dimensional models and forecast combinations.
		The predictive power of the two ratios appears strong during both recessions and expansions, stable over time, and consistent with well-established macro-finance theory.		
		
		\vspace{0.5cm}
		

	\end{abstract}

	\pagebreak

	\section{Introduction}
	
	Following the Global Financial Crisis and the Great Recession of 2008-2009, the interactions between asset prices, households' and firms' balance sheets, and real economic activity have become increasingly important in macroeconomic analysis (see, e.g., \citealt{GertlerGilchrist2018}; \citealt{MianSufi2018}, among many others).
	These interactions give rise to boom-and-bust financial cycles (\citealt{Borio2014}; \citealt{ClaessensKoseTerrones2011}; \citealt{DrehmannBorioTsatsaronis2012}), with major repercussions on business cycles.
	In particular, house prices and credit have received a great deal of attention and have been argued to provide ``the most parsimonious description of the financial cycle" (\citealt{Borio2014}).
	Empirical research has delivered substantial results on the predictive potential of financial-cycle indicators for business cycles (e.g., \citealt{MianSufiVerner2017}; \citealt{JordaSchularickTaylor2016}; \citealt{AdrianBoyarchenkoGiannone2019}). 
	However, comprehensive comparative evaluations are still needed to determine whether any specific variables, among those related to credit, balance sheets and housing, stand out as effective predictors of economic activity.
	
	This paper offers novel evidence on the ability of specific financial-cycle indicators to predict GDP over the medium term in the United States.
	Based on an extensive evaluation using a dataset of 262 quarterly macroeconomic and financial variables for the period 1960-2017, we find that two \textit{ratios} provide the best in-sample and out-of-sample predictions of GDP growth over periods of 1 to 5 years: 
	a cyclically-adjusted house price-to-rent (CAPR) ratio, calculated over the aggregate stock of owner-occupied housing (\citealt{DavisLehnertMartin2008}; \citealt{ContessiKerdnunvong2015}), 
	and the nonfinancial noncorporate business sector liabilities-to-income (NNBLI) ratio. 
	The CAPR ratio is a robust valuation metric for the housing market, representing the housing counterpart of the popular cyclically-adjusted price-to-earnings (CAPE) ratio introduced by \citet{CampbellShiller1998} for the stock market, while the NNBLI ratio measures the debt burden of noncorporate (small) businesses, which represent the vast majority (more than 80\% in recent years) of firms in the United States and account for almost 20\% of total revenues and value added.\footnote{Sources: U.S. Internal Revenue Service, \url{https://www.irs.gov/statistics/soi-tax-stats-integrated-business-data}; U.S. Bureau of Economic Analysis, \url{https://www.bea.gov/data/gdp/gross-domestic-product}.}
	To our knowledge, this is the first paper that shows the outstanding predictive power of these ratios.
	 
\begin{figure}[H]
		\caption{CAPR ratio and 5-year GDP growth}
		\vspace{-10pt}
		\centering
		\includegraphics[scale=0.22]{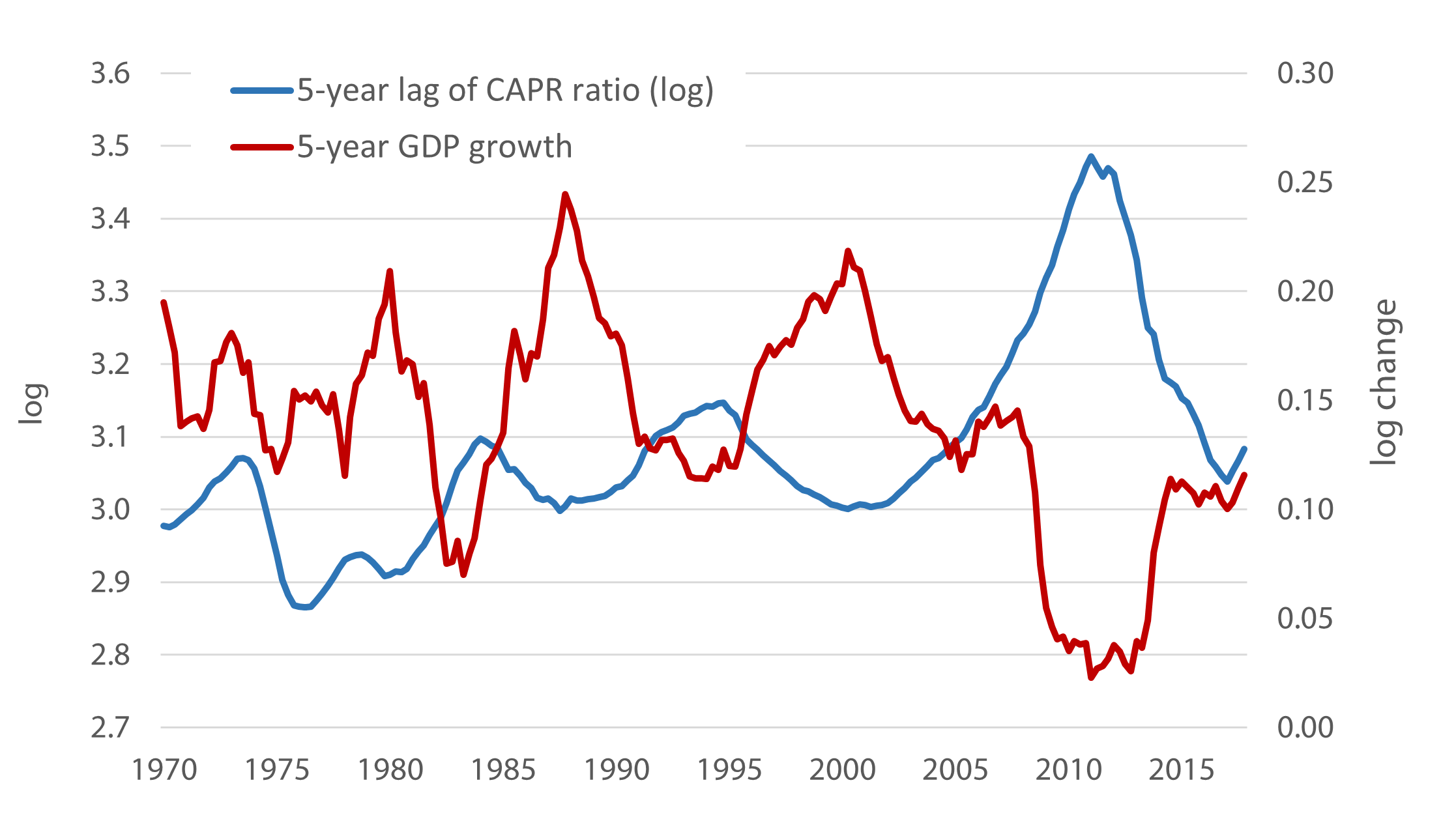}
		\subcaption*{\textit{Notes:} 
			The figure shows the 20-quarter lag of the log cyclically-adjusted house price-to-rent (CAPR) ratio (left axis) and the cumulative growth rate of GDP over the previous 20 quarters (right axis) in the United States, from 1970Q1 to 2017Q4. House prices and rents are measured by \citet{DavisLehnertMartin2008}, the data are available at \url{https://www.aei.org/historical-land-price-indicators/}. 
		}
		\label{fig:lcapr}
	\end{figure}

\begin{figure}[H]
		\caption{NNBLI ratio and 3-year GDP growth}
		\vspace{-10pt}
		\centering
		\includegraphics[scale=0.22]{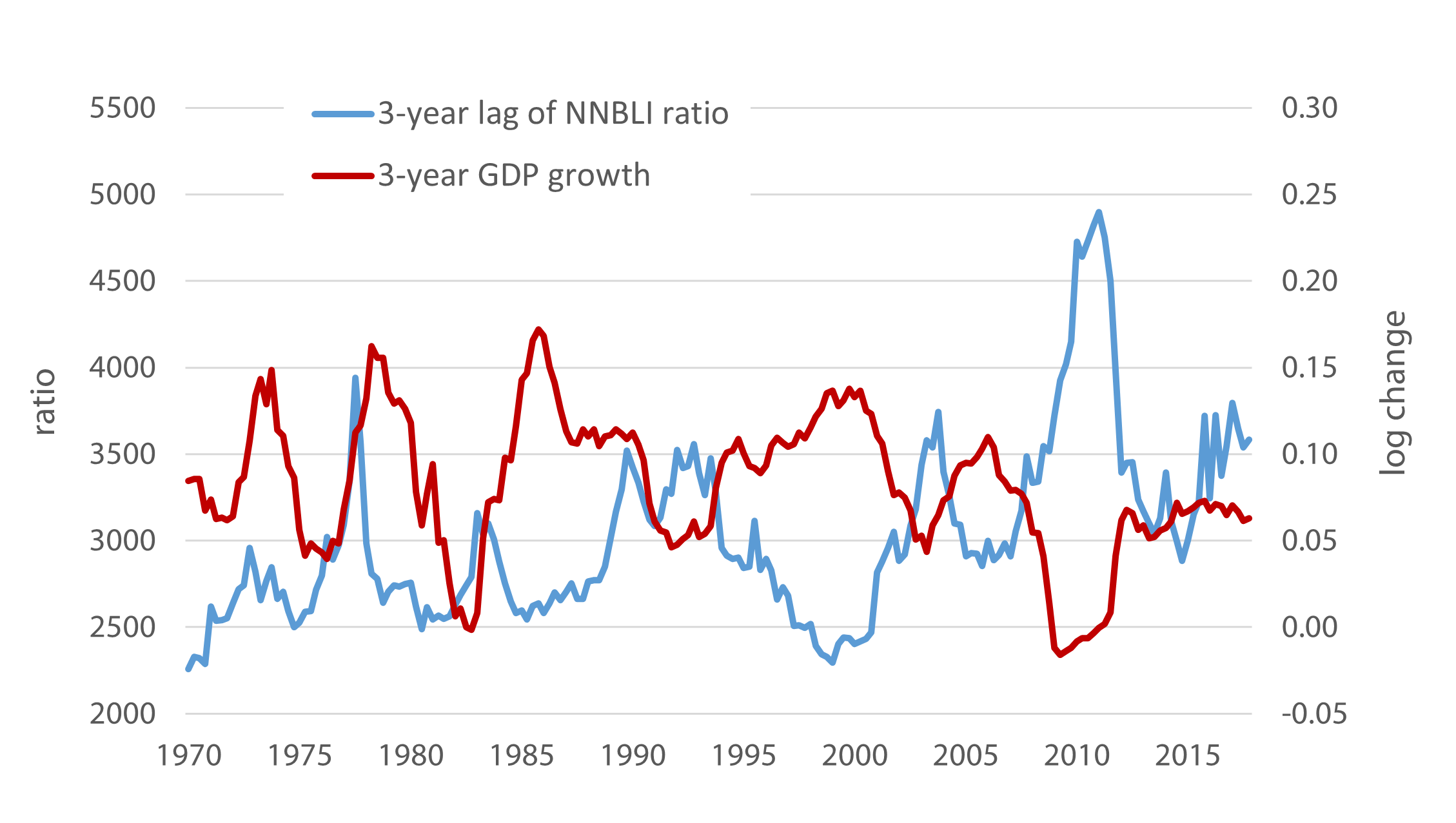}
		\subcaption*{\textit{Notes:}
			The figure shows the 12-quarter lag of the ratio of nonfinancial noncorporate business sector liabilities to disposable business income (NNBLI) (divided by 100, left axis) and the cumulative growth rate of GDP over the previous 12 quarters (right axis) in the United States, from 1970Q1 to 2017Q4.
			The data are from the FRED-QD dataset (\citealt{McCrackenNgFREDQD}).}
		\label{fig:nnbli}
	\end{figure}

	Both the CAPR and the NNBLI ratios are inversely related with subsequent medium-term economic activity. CAPR appears especially effective in predicting GDP growth over 3-5 years, while NNBLI provides its best predictions over horizons of 1-3 years. Figure \ref{fig:lcapr} plots the cumulative growth rate of GDP relative to 20 quarters before and compares it with the (log) CAPR ratio lagged by 20 quarters, while Figure \ref{fig:nnbli} compares the cumulative growth rate of GDP over the last 12 quarters with the 12-quarter lag of the NNBLI ratio. Both figures show a strong negative correlation (-0.7 and -0.61, respectively). 

	We first provide a set of baseline results, based on both in-sample and out-of-sample evaluation, using both direct forecasts produced by autoregressive distributed lag (ARDL) models and iterated forecasts by vector autoregressive (VAR) model. 
	These results offer unambiguous evidence of the special importance of CAPR and NNBLI, compared to all other predictors, and formalize the intuition provided by Figures \ref{fig:lcapr} and \ref{fig:nnbli}.
	We also find that forecasts produced by one-predictor models using the best financial-cycle ratios outperform forecasts by high-dimensional models and forecast combinations.
	This appears remarkable, since a large literature has shown that small forecasting models are often outperformed by models and methods that exploit a large amount of information, such as large Bayesian VARs (see, e.g., \citealt{BańburaGiannoneReichlin2010}; \citealt{CarrieroClarkMarcellino2015};  \citealt{Koop2013}; \citealt{KoopKorobilis2013}); 
	factor models (e.g., \citealt{StockWatson2002, StockWatson2011}; \citealt{ForniHallinLippiReichlin2001}),
	and forecast combinations (e.g., \citealt{StockWatson2003, StockWatson2004}).
	
	Next, we present a variety of extensions and robustness checks, including 
	the estimation of quantile regressions for GDP growth, 
	several checks on forecast instabilities, 
	forecasts by time-varying-parameter (TVP) models, 
	a comparison of alternative variable-selection methods, 
	and the evaluation of forecasts based on real-time (unrevised) data vintages, when available. 
	This extensive analysis provides several other striking results, which can be summarized as follows:
	the predictive content of the two ratios is not limited to recessions or periods of financial distress but is much more general. 
	First, quantile regressions reveal that their strong relationship with GDP growth over the medium term is stable across different parts of the GDP growth distribution, particularly in the case of CAPR. 
	This distinguishes the two ratios from general indicators of financial conditions, such as the National Financial Conditions Index (NFCI), which tend to be good predictors of the left tail of GDP, but are not very useful for the rest of the distribution, as shown by the recent literature on ``growth-at-risk" (\citealt{AdrianBoyarchenkoGiannone2019}; \citealt{AdrianGrinbergLiangMalikYu2022}) and confirmed by our results. 
	Also, while economic forecasts are generally found to be affected by substantial instabilities (e.g., \citealt{Rossi2021}, \citealt{StockWatson2003}, \citealt{ClementsHendry2006}), the predictive power of CAPR and NNBLI appears quite stable over time. In particular, the two ratios are shown to be top predictors before, during, and after the Global Financial Crisis. 

	The paper is related to several strands of the literature.
	First, it follows other comparative evaluations of predictors of economic activity based on large datasets (e.g., \citealt{StockWatson2003}; \citealt{BanerjeeMarcellinoMasten2005}; \citealt{MarcellinoStockWatson2003}).
	A large number of papers have focused on the role of financial variables, finding mixed evidence on their ability to forecast GDP (see, e.g., \citealt{StockWatson2003}; \citealt{ClaessensKose2017}). Some variables, such as the term spread of interest rates, exhibit good forecast performance but only in specific periods (\citealt{StockWatson2003};  \citealt{ChauvetPotter2013}).
	As already mentioned, the growth-at-risk literature (\citealt{AdrianBoyarchenkoGiannone2019}; \citealt{AdrianGrinbergLiangMalikYu2022}) shows that aggregate financial conditions help forecast tail risks to GDP.
	More specifically, our paper contributes to the literature on housing and credit cycles, and their predictive relationships with the business cycle.
	\citet{Leamer2007,Leamer2015} claims that ``housing \textit{is} the business cycle”, showing that it is the economic sector with the largest contribution to U.S. recessions and offers important early warnings. 
	Housing variables, such as building permits and housing starts, have long been used as leading indicators of GDP (\citealt{Green1997}; \citealt{CoulsonKim2000}).
	Also, there is extensive evidence on the prominent role of housing wealth shocks in the Great Recession and the subsequent slow recovery (e.g., \citealt{MianSufi2014}; \citealt{MianRaoSufi2013}). 
	Credit booms, especially those driven by mortgage credit, have been followed by deep recessions, slow recoveries and financial turmoil in recent decades (\citealt{JordàSchularickTaylor2013,JordaSchularickTaylor2016,JordaSchularickTaylor2017}; \citealt{MianSufi2018}). 
	In particular, high credit-to-GDP and household debt-to-GDP ratios predict lower GDP growth in the medium run (\citealt{MianSufiVerner2017}) and financial instability (\citealt{BorioLowe2002}). 
	Nonfinancial leverage has been a good predictor of growth vulnerability during the Great Recession (\citealt{ReichlinRiccoHasenzagl2019}). 
	\citet{AdrianGrinbergLiangMalikYu2022} show that loose financial conditions increase downside risks to GDP especially over medium-term horizons (1-3 years) and when credit-to-GDP growth is rapid. 
	Finally, the results of the paper on the strong predictive ability of the CAPR and NNBLI ratios appear to be consistent with theoretical frameworks in which credit market conditions and (housing-based) collateralized borrowing by firms and households are major drivers of economic fluctuations, including the popular ``financial accelerator" model by \citet{BernankeGertlerGilchrist1999} and more recent macro-finance models that explicitly incorporate the housing sector (e.g., \citealt{FavilukisLudvigsonVanNieuwerburgh2017}). At the end of the paper, we discuss in more detail the connections between our empirical findings and the theoretical insights provided by macro-finance research.

	The remainder of the paper is organized as follows:
	Section \ref{sec:data} introduces the data,
	Section \ref{sec:baseline} presents the baseline results of in-sample and (pseudo-)out-of-sample evaluation,
	Section \ref{sec:robustness} presents the extensions and robustness checks,
	Section \ref{sec:theory} discusses the macro-finance models that help explain the predictive power of CAPR and NNBLI,
	Section \ref{sec:conclusions} concludes.

	
	\vspace{-5pt}
	

	\section{Variables and data}\label{sec:data}
	
	We conduct our analysis using a large quarterly dataset (262 macroeconomic and financial variables) over the period 1960Q1-2017Q4. 
	The dataset combines the FRED-QD dataset\footnote{Available at \url{https://research.stlouisfed.org/econ/mccracken/fred-databases}. The FRED-QD dataset was officially launched in May 2018. This paper uses the 2018-06 vintage and the associated labels for variables. In section \ref{subsec:realtime}, we present results using historical (real-time) data vintages for a subset of variables for which they are available.} with the variables reported in Table \ref{tab:redux}, which include a variety of financial-cycle indicators (total credit to the non-financial sector, the credit-GDP ratio, households' mortgage debt and mortgage-income ratio, real house price growth and price-rent ratios, the ratio of residential fixed investment to GDP, real stock market prices and the cyclically-adjusted price-earnings ratio, and households' interest payments as a fraction of disposable income), as well as leading indicators (the OECD composite leading indicator and business confidence index) and the comprehensive National Financial Conditions Index (NFCI) by the Chicago Fed capturing a variety of other financial factors. Table \ref{tab:redux} also indicates the data transformations used for these additional variables.  
	The FRED-QD dataset is made of 248 variables, which cover in detail a wide spectrum of macro areas. \citet{McCrackenNgFREDQD} classify the variables into 14 groups: National Income and Product Accounts (NIPA); Industrial Production; Employment and Unemployment; Housing; Inventories, Orders, and Sales; Prices; Earnings and Productivity; Interest Rates; Money and Credit; Household Balance Sheets; Exchange Rates; Stock Markets; Non-Household Balance Sheets; Other. 
	For all variables in FRED-QD, we use the data transformations suggested by \citet{McCrackenNgFREDQD}, the only adaptation being the use of year-on-year instead of quarter-on-quarter changes/growth rates. In addition, for all rates and ratios (interest rates and spreads, unemployment rates, exchange rates, balance-sheet ratios, etc.) for which differencing is suggested, we also consider the levels, i.e., non-transformed data.
	Table \ref{tab:fredq_legend} lists the subset of FRED-QD variables that will be reported among the best predictors of GDP in the baseline results.
	In all results, uppercase labels will identify variables from FRED-QD and lowercase labels the additional set of variables, as shown in Tables \ref{tab:redux} and \ref{tab:fredq_legend}.

	In analogy with the popular cyclically-adjusted price-to-earnings ratio (CAPE) proposed by \citet{CampbellShiller1998} for the stock market, which is calculated by dividing the real S\&P 500 index by the 10-year moving average of real earnings on the index, 
	the CAPR ratio is calculated by dividing real house prices by the 10-year average real rents:
	\[
	CAPR_t = \frac{HPI_t}{\frac{1}{40}\sum_{i=1}^{40}R_{t-1}} 
	\]
	where $HPI_t$ denotes the house price index in real terms at time $t$ (in quarters) and $R_t$ is the imputed real rent at time $t$.
	Following \citet{ContessiKerdnunvong2015}, we calculate the CAPR using data by \citet{DavisLehnertMartin2008}, who  measure house prices and (imputed) rents on the entire stock of owner-occupied houses, representing the large majority of houses in the United States.\footnote{See \citet{DavisLehnertMartin2008} for methodological details. 
		The raw data on house prices and rents over the period 1960Q1-2018Q2 are available at \url{https://www.aei.org/historical-land-price-indicators/}. 
		To calculate the 40-quarter moving average of rents from 1960Q1 to 1970Q1, we extrapolate the rent series backward from 1960Q1 to 1950Q1 using the growth rate of the rent of primary residence, measured by the U.S. Bureau of Labor Statistics (BLS) and provided by FRED at \url{https://fred.stlouisfed.org/}.
		The home-ownership rate in the United States in the period 1965-2017 ranged between 63\% and 69\% (see \url{https://www.census.gov/housing/hvs/index.html}).} 
	\citet{ContessiKerdnunvong2015} use the ratio to test for bubbles in the housing market.

	The NNBLI ratio is provided by the FRED-QD dataset and is calculated as the ratio of liabilities to business disposable income. The noncorporate business sector primarily consists of partnerships and sole proprietorships, which represent the typical organizational forms of small firms.\footnote{Noncorporate businesses are pass-through firms, i.e., their income is passed on to firms' owners and treated (taxed) as personal income.  As a result, by construction, the sector has zero disposable income in national accounts (see the Fed's Financial Accounts of the United States at \url{https://www.federalreserve.gov/releases/z1/} and the Integrated Macroeconomic Accounts by the U.S. Bureau of Economic Analysis at \url{https://www.bea.gov/data/special-topics/integrated-macroeconomic-accounts}.)
		To calculate business disposable income at the denominator of the NNBLI ratio, FRED-QD uses data on corporate cash flows and taxes, which provide a good proxy. }

		\pagebreak

	\newgeometry{left=1.5cm,top=1cm}
	\begin{landscape}
		
		\begin{table}[]
			\caption{Variables in addition to the FRED-QD dataset}%
			\small{
				\begin{tabular}{lllll}
					label & description                                                   & data source         & source label          & transformation       \\
					
					\hline
					bci         & OECD Business Confidence Index                                & OECD                & —                   & $x_t$               \\
					cli         & OECD Composite Leading Indicator                              & OECD                & —                   & $x_t$               \\
					sp500       & Real S\&P500 index growth                                     & Shiller             & —                 & $\ln(x_t)-\ln(x_{t-4})$ \\
					cape        & Cyclically adjusted price/earnings ratio 						& Shiller             & —                   & $\ln(x_t)$           \\
					cred        & Real credit growth                                            & FRED                & \uppercase{crdqusapabis/cpiaucsl} & $\ln(x_t)-\ln(x_{t-4})$ \\
					cred\_gdp   & Credit/GDP ratio                                            & BIS                 & —            & $x_t$               \\
					
					hpi         & Real house price growth                                       & Shiller             & —  & $\ln(x_t)-\ln(x_{t-4})$ \\
					pr          & Price/rent ratio                                              & AEI & —                     & $\ln(x_t)$           \\
					capr        & Cyclically-adjusted price/rent ratio   & AEI; FRED & —$\quad$ ; CUUR0000SEHA, CPIAUCNS                  & $\ln(x_t)$           \\
					mortg       & Households' real mortgage debt growth                                    & FRED                & \uppercase{hhmsdodns/cpiaucsl}    & $\ln(x_t)-\ln(x_{t-4})$ \\
					mortg\_inc  & Households' mortgage/income ratio                                         & FRED                & \uppercase{hhmsdodns/dspi}        & $x_t$               \\
					prfi\_gdp   & Private residential fixed investment/GDP ratio                & FRED                & \uppercase{prfi/gdp}              & $x_t$               \\
					pip\_inc    & Personal interest payments/income ratio                       & FRED                & \uppercase{b069rc1/dspi}          & $x_t$               \\
					
					nfci        & Chicago Fed national financial condition index                & FRED                & \uppercase{ncfi}                  & $x_t$              \\
					\hline
				\end{tabular}
			\hfill*
			}
			\subcaption*{\textit{Notes:} Shiller = http://www.econ.yale.edu/\textasciitilde{}shiller/data.htm; AEI = American Enterprise Institute, https://www.aei.org/historical-land-price-indicators/, based on original data by Davis et al. (2008) (using the Case-Shiller price index after 2000).} 
			\label{tab:redux}
		\end{table}
		
	\end{landscape}	
	\restoregeometry
	
	\begin{landscape}
		\begin{table}[H]
			\caption{Best performing FRED-QD variables}
			\centering
			{
				\scalebox{0.84}[0.71]{
					\begin{tabular}{ll}
						label & description	\\
						\hline
						AAA             & Moody’s Seasoned Aaa Corporate Bond Yield© (Percent)                                                                       \\
						AHETPIx         & Real Average Hourly Earnings of Production and Nonsupervisory Employees: Total                                             \\
						AMDMUOx         & Real Value of Manufacturers’ Unfilled Orders for Durable Goods Industries (Millions of 2012 Dollars), deflated by Core PCE \\
						B021RE1Q156NBEA & Shares of gross domestic product: Imports of goods and services (Percent)                                                  \\
						CONSPIx         & Nonrevolving consumer credit to Personal Income                                                                            \\
						DOTSRG3Q086SBEA & Personal consumption expenditures: Other services (chain-type price index)                                                 \\
						FEDFUNDS		& Effective Federal Funds Rate (Percent)																					 \\
						GS1TB3Mx        & 1-Year Treasury Constant Maturity Minus 3-Month Treasury Bill, secondary market                                            \\
						HNOREMQ027Sx    & Real Real Estate Assets of Households and Nonprofit Organizations (Billions of 2012 Dollars), deflated by Core PCE         \\
						IPCONGD         & Industrial Production: Consumer Goods (Index 2012=100)                                                                     \\
						IPNCONGD        & Industrial Production: Nondurable Consumer Goods (Index 2012=100)                                                          \\
						ISRATIOx        & Total Business: Inventories to Sales Ratio                                                                                 \\
						LIABPIx         & Liabilities of Households and Nonprofit Organizations Relative to Personal Disposable Income (Percent)                     \\
						MORTGAGE30US    & 30-Year Conventional Mortgage Rate© (Percent)                                                                              \\
						NNBTILQ027SBDIx & Nonfinancial Noncorporate Business Sector Liabilities to Disposable Business Income (Percent)                              \\
						NWPIx           & Net Worth of Households and Nonprofit Organizations Relative to Disposable Personal Income (Percent)                       \\
						PCECC96         & Real Personal Consumption Expenditures (Billions of Chained 2009 Dollars)                                                  \\
						PRFIx           & Real private fixed investment: Residential (Billions of Chained 2009 Dollars), deflated using PCE                          \\
						REVOLSLx        & Total Real Revolving Credit Owned and Securitized, Outstanding (Billions of 2012 Dollars), deflated by Core PCE            \\
						S\&P: div yield & S\&P’s Composite Common Stock: Dividend Yield                                                                              \\
						TLBSNNCBBDIx    & Nonfinancial Corporate Business Sector Liabilities to Disposable Business Income (Percent)                                 \\
						UNRATESTx       & Unemployment Rate less than 27 weeks (Percent)                                                                             \\
						USMINE          & All Employees: Mining and logging (Thousands of Persons) \\
						\hline                                                                  
					\end{tabular}%
				\hfill*
				}
				\subcaption*{\textit{Notes:} The table reports the subset of variables from the FRED-QD dataset (\citealt{McCrackenNgFREDQD}) that rank among the best predictors of GDP in this paper (see Sections \ref{subsec:is} and \ref{subsec:oos}.) See Section \ref{sec:data} for more details on data transformations. Starting in the FRED-QD vintage 2018-09, the label NNBTILQ027SBDIx has been replaced with TLBSNNBBDIx.}
			}
			\label{tab:fredq_legend} 
		\end{table}
		
	\end{landscape}
	
	\pagebreak
	
	\restoregeometry

	\section{Baseline results}\label{sec:baseline}
	
	\subsection{In-sample evaluation}\label{subsec:is}

	Following \citet{StockWatson2003}, we conduct our in-sample evaluation of predictive power using autoregressive distributed lag (ARDL) models for multi-period cumulative GDP growth rates. 
	More specifically, we predict the $h$-quarter-ahead cumulative GDP growth rate, with $h=4,12,20$, using a bivariate model that includes one predictor at a time plus lags of GDP growth. We also consider results for $h=28$, which corroborate those obtained for $h=20$ and are not reported in the interest of space.
	The economic significance of alternative predictors is then evaluated using the $R^2$ of the regressions.
	
	The ARDL regressions are as follows: 
	\begin{equation}\label{eq:ardl}
		y_t^{h} = \beta_0 +  \beta_{1}(L) y_{t-h} +  \beta_{2,i} (L) x_{i, t-h}+ u_t
	\end{equation}
	where $y_t^h$ is the log approximation of the cumulative GDP growth rate over a period of length $h$, i.e.,  $y_t^h = \ln(GDP_t)-\ln(GDP_{t-h}) $, $y_t$ is the log approximation of the year-on-year GDP growth rate at time $t$, i.e., $y_t= \ln(GDP_t)-\ln(GDP_{t-4})$, $x_{i,t}$ is the $i$-th candidate predictor, $u_t$ is the error term,  $\beta_0$ is a constant, $\beta_{1}(L)$ and $\beta_{2,i}(L)$ are lag polynomials, such that $\beta_{1}(L) y_{t-h} = \sum_{j=1}^{p} \beta_{1,j} y_{t-h+1-j}  $ and $\beta_{2,i}(L) x_{i, t-h} = \sum_{j=1}^{q} \beta_{2,i,j} x_{i, t-h+1-j}  $. 
	
	The models are estimated using data in the time range 1974Q1-2017Q4. The starting date is chosen so as to ensure that 95\% of the time series have no missing data in the sample (shorter series are excluded here but will be used in the out-of-sample evaluation of section \ref{subsec:oos}). 
	To ensure perfect comparability of $R^2$, we fix the lag length across models. In particular, both $p$ and $q$ are set to 5 to adequately account for serial correlation in quarterly time series.
	
	Table \ref{tab:r2} reports the $R^2$ of the regressions.  Predictors are listed in descending order of $R^2$ and only the best 10 are reported for each value of $h$, out of the 262 predictors in the dataset. The cyclically-adjusted log price-rent ratio (label: $capr$) dominates over longer horizons. It is the best predictor for $h=12,20$ and the second best for $h=4$. The OECD composite leading indicator ($cli$) is a particularly strong predictor over short horizons. It ranks first for $h=4$, then it gradually falls down the ranking as the horizon increases (5th for $h=12$, 9th for $h=20$).
	Most of the top positions are occupied by financial-cycle indicators.
	The noncorporate liabilities-income ratio or NNBLI (FRED-QD label: \textit{NNBTILQ027SBDIx}\footnote{Starting in the FRED-QD vintage 2018-09, this label has been replaced with \textit{TLBSNNBBDIx}.}) is the third-best predictor for $h = 4,12$, after CAPR and the unadjusted price-rent ratio ($pr$), and the fifth best predictor for $h=20$.
	Private residential fixed investment, both as a share of GDP and in growth rates (labels: \textit{prfi\_gdp} and \textit{PRFIx}, respectively), performs especially well over the 4-quarter horizon, while the mortgage/income ratio (\textit{mortg\_inc}), households' liabilities and net worth relative to disposable income (\textit{LIABPIx} and \textit{NWPIx}), and the credit/GDP ratio (\textit{cred\_gdp}) are effective predictors over longer horizons.

	Among the other variables, the unfilled orders for nondurable goods (\textit{AMDMUOx}) feature in the top 10 for $h=4,12$, while the inventories to sales ratio (\textit{ISRATIOx}) for $h=12,20$.
	
	The usefulness of the CAPR appears even more remarkable if the lag orders $p$ and $q$ are selected by the Bayes information criterion (BIC) (we first select $p$, then $q$ conditional on $p$, given the maximum lag length of 5). In this case, it is the best predictor at all horizons (results are not reported).

	Finally, the results on CAPR are not simply determined by the specific indices of house prices and rents considered. To check this, we replicate the analysis on two alternative measures of CAPR, one calculated using the house price index by \citet{Shiller2015} and the average rent of primary residence in U.S. cities by the Bureau of Labor Statistics, the second using the nominal house price and rent indices by the OECD. 
	Taking these two measures, CAPR still ranks among the top predictors for all horizons. However, the best results are obtained using the data on the aggregate stock of owner-occupied housing by \citet{DavisLehnertMartin2008}.

	\begin{table}[H]
		\centering
		\caption{Regression $R^2$ of single-predictor ARDL models}%
		\scalebox{0.8}[0.8]{
			\begin{tabular}{l|ll|ll|ll}
				\hline \hline 
				& \multicolumn{2}{c|}{$h=4$}       & \multicolumn{2}{c|}{$h=12$} & \multicolumn{2}{c}{$h=20$}  \\
				\hline
				1  & cli               & 0.43 & capr            & 0.63 & capr            & 0.68 \\
				2  & capr              & 0.40 & pr              & 0.57 & pr              & 0.65 \\
				3  & NNBTILQ027SBDIx   & 0.36 & NNBTILQ027SBDIx & 0.56 & LIABPIx         & 0.58 \\
				4  & prfi\_gdp         & 0.35 & AMDMUOx         & 0.55 & mortg\_inc      & 0.58 \\
				5  & pr                & 0.35 & cli             & 0.50 & NNBTILQ027SBDIx & 0.55 \\
				6  & DOTSRG3Q086SBEA   & 0.33 & mortg\_inc      & 0.45 & ISRATIOx        & 0.53 \\
				7  & nfci              & 0.33 & LIABPIx         & 0.44 & cred\_gdp       & 0.53 \\
				8  & PRFIx             & 0.33 & ISRATIOx        & 0.44 & NWPIx           & 0.47 \\
				9  & AMDMUOx           & 0.32 & cred\_gdp       & 0.42 & AAA             & 0.47 \\
				10 & IPCONGD           & 0.31 & NWPIx           & 0.41 & CONSPIx         & 0.44 \\
				\hline \hline 
			\end{tabular}
		}
		\subcaption*{\textit{Notes:} For each $h$, the dependent variable is GDP growth over $h$ quarters (cumulative). Models are estimated using from 1974Q1 to 2017Q4. Please refer to Tables \ref{tab:redux} and \ref{tab:fredq_legend} for a description of the variables.}
		\label{tab:r2}
	\end{table}

	\subsection{Out-of-sample evaluation}\label{subsec:oos}

	The second part of the analysis evaluates the forecasting power of the predictors out of the estimation sample.   
	To this aim, we track direct and iterated forecasts over time using a recursive-window scheme.\footnote{As explained in section \ref{subsec:rolling}, we also consider rolling windows of various fixed lengths. However, the best forecasts are generally achieved in the recursive-window case, so we use this for our baseline results.} Direct forecasts are made using multi-period ARDL models, while iterated forecasts are produced by bivariate VAR models. The forecast horizons are the same as in the in-sample evaluation, i.e., 4, 12 and 20 quarters (again, $h=28$ is considered but not reported as it only strengthens the main results for $h=20$). Time series of forecasts over different horizons are constructed for each one of the competing models and then used for comparison. In particular, predictors are evaluated using the mean square forecast errors (MSFE) computed over the period 1990Q1-2017Q4. 
	Given the maximum forecast horizon of 20 quarters, this implies setting the ending point of the shortest estimation window to 1985Q1, which is accommodated by moving the starting point to an earlier date than in the previous sections, namely 1968Q2.\footnote{As the maximum number of lags is set to 5, this specific start date is chosen to ensure that all the VAR models are estimated using data as far back as 1967Q1, which is the first quarter in which data are available (after transformation) for at least 90\% of the time series in the dataset. Concerning the multi-step ARDL models, the range of data used for estimation depends on the relevant horizon (in the case $h=28$, estimation uses data as far back as 1960Q1). Also, given the unbalanced nature of the dataset, the actual starting date of the sample will adjust to the availability of the time series used for estimation. In sections \ref{subsubsec:oos_ardl} and \ref{subsubsec:oos_var}, which consider one predictor at a time, all predictors are used regardless of the length of the respective time series. Therefore, when interpreting the results it should be kept in mind that a fraction of the predictors have fewer observations available for estimation than others. To ensure that the MSFE is computed over the same timespan for all predictors, only variables that have sufficient data to produce forecasts for 1990Q1 should be considered. However, even if the variables with an insufficient number of observations are included in the evaluation, none of them ranks among the best-performing predictors reported in sections \ref{subsubsec:oos_ardl} and \ref{subsubsec:oos_var}.
		The high-dimensional models presented in sections \ref{subsubsec:hd_models} exclude from estimation those variables (21 in total) whose time series start after 1967Q1 (as they would lead to discard observations for all other regressors), while the 14 predictors with the shortest time series (which do not have enough data to make direct forecasts for as early as 1990Q1, at least over the longest horizon) are excluded from forecast combinations.\label{fn_exclusions}}

	Competing models are initially estimated on the shortest sample 1968Q2-1985Q1 and used to make forecasts for the period from 1986Q1 (4-quarter-ahead forecast) to 1990Q1 (20-quarter-ahead forecast). Then the sample is recursively expanded by one quarter at a time and the estimation and forecasting steps are repeated in each iteration (e.g., in the second iteration models are estimated on the sample 1968Q2-1985Q2 and then used to produce forecasts for the period 1986Q2-1990Q2).

	\subsubsection{Direct forecasts: ARDL models}\label{subsubsec:oos_ardl}
	The first out-of-sample evaluation procedure is based on direct forecasts produced by bivariate ARDL models as in \eqref{eq:ardl}, in which the lag lenghts $p$ and $q$ are selected recursively using the BIC.  
	Let $ \widehat{y}_{i,t+h|t}^{h}$ denote the direct out-of-sample forecasts of $ y_{t+h}^{h}$ made by the model with the $i$-th predictor, estimated on data up to time $t$:
	\begin{equation}\label{eq:ardl_hat}
		\widehat{y}_{i,t+h|t}^{h} = \widehat{\beta}_0^{(t)} +  \widehat{\beta}_{1}^{(t)}(L) y_{t} +  \widehat{\beta}_{2,i}^{(t)} (L) x_{i,t}
	\end{equation}
	and let $\widehat{u}_{i,t+h|t}=  y_{t+h}^{h} -  \widehat{y}_{i,t+h|t}^{h} $ denote the forecast error incurred by the model at time $t+h$.
	Each $i$-th predictor is ranked based on the following MSFE:
	\begin{equation}
		MSFE_{i,h} = \frac{1}{T_1 -h -T_0 +1}\sum_{t=T_0}^{T_1-h} \left( \widehat{u}_{i,t+h|t} \right)^{2}
	\end{equation}
	where $T_0$ is the end date of the shortest sample and $T_1-h$ is the end date of the longest sample.
	
	Table \ref{tab:mse_ardl} reports the MSFE of the ARDL models relative to an AR model, used as a benchmark.\footnote{For the AR model, we consider both direct forecasts, i.e., produced by model \eqref{eq:ardl} without the terms associated with $x_i$ (and with the lag length $p$ selected recursively by the BIC), and iterated forecasts, i.e., using specification \eqref{eq:var} below. 
	Since iterated forecasts are generally more accurate, we use these as our benchmark to calculate the relative MSFE of all models.	
	Thus, the MSFE of ARDL (direct) forecasts and VAR (iterated) forecasts are fully comparable throughout the paper.
		The root mean square forecast error of the benchmark AR is: 0.0174 for $h=4$,  0.0397 for $h=12$ and 0.0561 for $h=20$.} 
	The log CAPR ratio is by far the best predictor over long horizons ($h=20,28$) and the second-best predictor for both $h=4$ and $h=12$. The NNBLI ratio is the most effective predictor for $h =4,12$ and the second-best predictor at a 5-year horizon. 
	Forecast gains over the benchmark are substantial: the log CAPR ratio and the NNBLI ratio achieve MSFE values as low as 0.35 (log CAPR for $h=20$) and 0.40 (NNBLI for $h=12$). For $h=4$, NNBLI has a relative MSFE of 0.55. Such results are all the more remarkable if one considers that only two variables have MSFE values lower than 0.8 for $h=4$, and only three variables for $h=12$ and $h=28$.
	The absolute root mean square forecast error (RMSFE) also helps appreciate the forecast performance of CAPR and NNBLI. In terms of 20-quarter cumulative GDP growth, the RMSFE of CAPR is  3.32\%, corresponding to an average annual error of 0.66 percentage points of GDP growth for 5 years. For $h=12$, the RMSFE of NNBLI is 2.50\%, corresponding to an average annual error of 0.83\%.
	
	Other top performers include the unfilled orders for nondurable goods (\textit{AMDMUOx}) (for $h=4,12$), the OECD composite leading indicator ($cli$) (for $h=4,12,20$), the unadjusted price-rent ratio ($pr$) (for $h=4,20$), industrial production of consumer goods (\textit{IPCONGD}) (for $h=4,12,20$), the ratio of households' net worth to disposable income (\textit{NWPIx}) (for $h=4,12$) and the inventories to sales ratio (\textit{ISRATIOx}) (for $h=20$).

	\begin{table}[H]
		\centering
		\caption{Direct forecasts by single-predictor ARDL models: mean squared forecast errors}%
		\scalebox{0.8}[0.8]{
			\begin{tabular}{l|ll|ll|ll}
				\hline \hline 
				& \multicolumn{2}{c|}{$h=4$}       & \multicolumn{2}{c|}{$h=12$} & \multicolumn{2}{c}{$h=20$}  \\
				\hline
				1  & NNBTILQ027SBDIx & 0.55 & NNBTILQ027SBDIx & 0.40 & capr            & 0.35 \\
				2  & capr            & 0.74 & capr            & 0.63 & NNBTILQ027SBDIx & 0.53 \\
				3  & AMDMUOx         & 0.84 & cli             & 0.76 & pr              & 0.56 \\
				4  & pr              & 0.86 & AMDMUOx         & 0.95 & ISRATIOx        & 0.95 \\
				5  & cli             & 0.87 & NWPIx           & 1.01 & cli             & 0.96 \\
				6  & IPCONGD         & 0.88 & cred_gdp        & 1.05 & LIABPIx         & 1.06 \\
				7  & NWPIx           & 0.89 & UNRATESTx       & 1.05 & UNRATESTx       & 1.06 \\
				8  & IPNCONGD        & 0.89 & IPCONGD         & 1.06 & AHETPIx         & 1.10 \\
				9  & PCECC96         & 0.91 & fedfunds        & 1.07 & MORTGAGE30US    & 1.11 \\
				10 & prfi_gdp        & 0.91 & B021RE1Q156NBEA & 1.07 & IPCONGD         & 1.12\\
				\hline \hline 
			\end{tabular}
		}
		\subcaption*{\textit{Notes}: The table shows the mean squared forecast errors (MSFE) for the $h$-quarter (cumulative) GDP growth rate, relative to an AR model. All models are estimated on recursive windows (shortest sample 1968Q2-1985Q1, longest sample 1968Q2-2016Q4) and MSFE are calculated over the period 1990Q1-2017Q4. Please refer to Tables \ref{tab:redux} and \ref{tab:fredq_legend} for a description of the variables.}
		\label{tab:mse_ardl} 
	\end{table}

	\subsubsection{Iterated forecasts: VAR models}\label{subsubsec:oos_var}
	The second out-of-sample approach uses VAR models to compute multi-step-ahead iterated forecasts of GDP. 
	Evaluation of predictors is conducted using models that include only real GDP growth and one predictor at a time. The $i$-th VAR can be written as:
	\begin{equation}\label{eq:var}
		\widetilde{y}_t^{(i)} = a_0^{(i)} + \sum_{j=1}^{p} B_j^{(i)} \widetilde{y}_{t-j}^{(i)} + \varepsilon_t^{(i)}
	\end{equation}
	where $\widetilde{y}_t^{(i)}$ is the vector containing the year-on-year growth rate of real GDP and the $i$-th predictor at time $t$,  $a_0^{(i)}$ is a  $2 \times 1$ vector of constants, $B_j^{(i)}$ is a $2 \times 2$ matrix of coefficients, $\forall j=1,\dots,p$, and $\varepsilon_t^{(i)}$ is a  $2 \times 1$ vector of error terms.
	The lag length $p$ is recursively selected by the BIC for the whole system and the maximum length is again fixed at 5.
	
	Predictors are ranked based on the performance of the VARs in terms of forecasts of the $h$-period-ahead GDP level:
	\begin{equation}
		MSFE_{GDP}^{(i,h)} = \frac{1}{T_1-h-T_0+1}\sum_{t=T_0}^{T_1-h} \left( \ln(GDP_{t+h}) - \ln(\widehat{GDP}^{(i)}_{t+h|t}) \right)^{2}
	\end{equation}
	where $ \ln(\widehat{GDP}^{(i)}_{t+h|t})$ is the forecast of the log GDP level for period $t+h$ obtained from model $i$ by cumulating the growth rate forecasts over time.

	Table \ref{tab:mse_var} reports the MSFE of the best 10 VAR models for each horizon, relative to the benchmark AR.
	The top positions remain largely unchanged with respect to Table \ref{tab:mse_ardl}. Once again, the log CAPR is the best predictor for $h=20$ and ranks second for $h=4,12$. The NNBLI ratio is still the most useful predictor for $h=4,12$. 
	
	Just as in Table \ref{tab:mse_ardl}, the forecast gains provided by CAPR and NNBLI over the benchmark AR are substantial for every $h$. In particular, for $h=12$ the relative MSFE of NNBLI is 0.34, corresponding to an absolute RMSFE of 2.31\% (an annual average of 0.77 percentage points of GDP growth). For $h=20$, the log CAPR gives a relative MSFE of 0.40 and an absolute RMSFE of 3.56\% (annual average: 0.71\%). For $h=4$, the relative MSFE of NNBLI is 0.55, corresponding to an absolute RMSFE of 1.3\%.\footnote{We also check that the predictive gains provided by CAPR and NNBLI compared to the AR model are statistically significant at all horizons, using the ENC-F test by \citet{ClarkMcCracken2001}, which is suitable for comparing nested models.} 
	
	As for the other top predictors, once again private residential fixed investment (as a share of GDP) ranks highly over the shorter horizon of 4 quarters, while the OECD composite leading indicator ($cli$) and the unfilled orders for durable goods (\textit{AMDMUOx}) rank highly over all horizons.

	\begin{table}[H]
		\centering
		\caption{Iterated forecasts by bivariate VAR models: mean squared forecast errors}%
		\scalebox{0.8}[0.8]{
			\begin{tabular}{l|ll|ll|ll}
				\hline \hline 
				& \multicolumn{2}{c|}{$h=4$}       & \multicolumn{2}{c|}{$h=12$} & \multicolumn{2}{c}{$h=20$}   \\
				\hline
				1  & NNBTILQ027SBDIx & 0.55 & NNBTILQ027SBDIx & 0.34 & capr            & 0.40 \\
				2  & capr            & 0.71 & capr            & 0.53 & NNBTILQ027SBDIx & 0.48 \\
				3  & pr              & 0.76 & pr              & 0.69 & pr              & 0.53 \\
				4  & AMDMUOx         & 0.80 & cli             & 0.71 & HNOREMQ027Sx    & 0.80 \\
				5  & TLBSNNCBBDIx    & 0.83 & AMDMUOx         & 0.75 & AMDMUOx         & 0.81 \\
				6  & prfi\_gdp       & 0.84 & UNRATESTx       & 0.84 & UNRATESTx       & 0.81 \\
				7  & cli             & 0.87 & USMINE          & 0.86 & ISRATIOx        & 0.81 \\
				8  & S\&P: div. yield& 0.88 & IPNCONGD        & 0.86 & B021RE1Q156NBEA & 0.82 \\
				9  & IPCONGD         & 0.88 & B021RE1Q156NBEA & 0.87 & cli             & 0.83 \\
				10 & cape            & 0.89 & GS1TB3Mx        & 0.90 & REVOLSLx        & 0.84 \\
				\hline \hline 
			\end{tabular}
		}
		\subcaption*{\textit{Notes}: The table shows the mean squared forecast errors (MSFE) for for $h$-quarter-ahead GDP, relative to an AR model. All models are estimated on recursive windows (shortest sample 1968Q2-1985Q1, longest sample 1968Q2-2016Q4) and MSFE are calculated over the period 1990Q1-2017Q4. Please refer to Tables \ref{tab:redux} and \ref{tab:fredq_legend} for a description of the variables.}
		\label{tab:mse_var} 
	\end{table}

	\subsubsection{Comparison with high-dimensional forecasting models and forecast combinations}\label{subsubsec:hd_models}

\begin{figure}[h]
		\caption{High-dimensional models: sparsity in the LASSO VAR}
		\centering
		\includegraphics[scale=0.043]{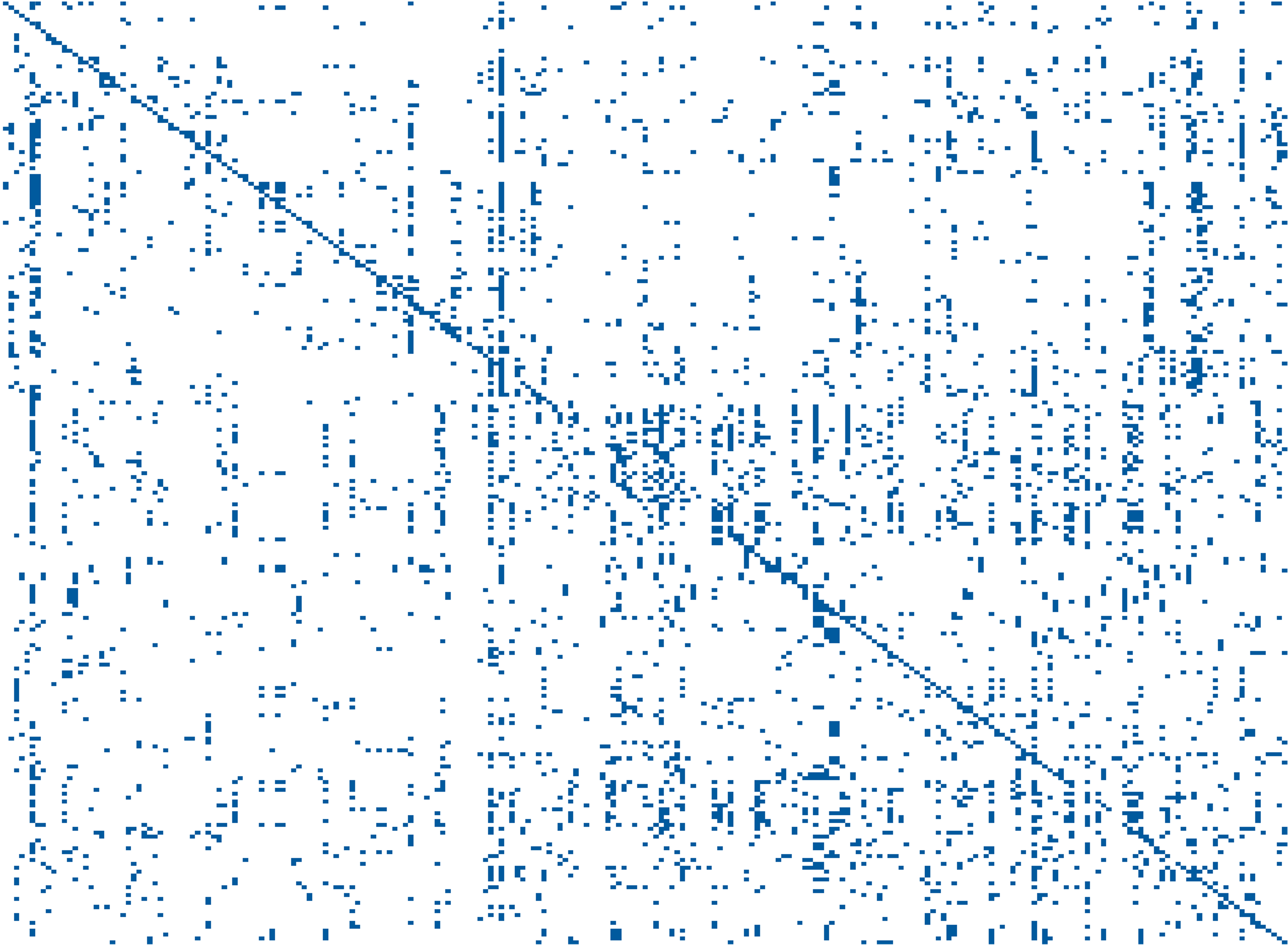} 
		\vspace{3pt}
		\subcaption*{\textit{Notes:} The figure provides a visual summary of variable selection in the LASSO VAR model, estimated over the sample 1968Q2-2016Q4. The figure depicts a table in which each row corresponds to an equation in the large VAR, and each column identifies a right-hand-side variable (predictor). Colored cells identify predictors with non-zero coefficients.}
		\label{fig:lassovar_selection}
	\end{figure}

	Finally, we assess whether the forecasts produced by the best single-predictor models are outperformed by those produced by methods that pool all available information. We consider three models that use all predictors at the estimation stage: (i) a Large Bayesian VAR (LBVAR) using the approach by \citet{BańburaGiannoneReichlin2010}, (ii) a LASSO VAR, i.e., a VAR estimated using a LASSO penalty (\citealt{HsuHungChang2008}), and (iii) a factor model using principal components (a VAR model for GDP growth and a set of principal components extracted from all predictors). 
	The three models reflect different approaches to the problem of high dimensionality. The LBVAR approach retains all the available predictors in the forecasting model, applying a shrinkage method that does not restrict any coefficient to be exactly zero. The LASSO VAR performs variable selection by setting a subset of coefficients exactly to zero. The principal component approach reduces the dimension of the model by summarizing the dataset of predictors into a small number of factors. 
	We estimate the LBVAR and LASSO VAR using grids of values for the shrinkage/penalty parameters and report the best results for each forecast horizon.\footnote{The LBVAR model is estimated using a grid of possible values for the shrinkage parameters $\lambda$ and $\tau$, based on notation by \citet{BańburaGiannoneReichlin2010}, where $\lambda$ is an inverse measure of the tightness of the prior on the VAR coefficients and  $\tau$ determines the tightness of the additional prior on the sum of coefficients. In particular, the values considered for $\lambda$ lie in the range [0.0001, 0.1], while $\tau$ can assume values 10 or 100. We then select the values that provide the best forecast performance. The priors on model parameters are set following \citet{AlessandriMumtaz2017}. Forecasts are produced using the posterior means of the parameters.
	As for the LASSO VAR, using $\lambda$ to denote the penalty parameter of the LASSO estimator, we consider $\lambda=0.00025, 0.0005, 0.00075, 0.001, 0.00125, 0.0015 $. Again, we select the value that provides the best forecast performance.
	The maximum lag length is 5 quarters for the BVAR and the factor model. To prevent computational problems, we consider a lag length of 1 for the LASSO VAR.
	}
	As for the factor VAR, for each horizon we select the number of factors (between 1 and 6) that gives the best forecasts. For all models, we produce iterated forecasts of GDP using the same recursive-window scheme as in previous sections.
	As in equation \eqref{eq:var}, we include the 4-quarter growth rate of GDP in the models.
	To get a sense of the degree of sparsity in the LASSO VAR model, Figure \ref{fig:lassovar_selection} depicts the regressors selected by LASSO (in the columns) for all equations of the model (in the rows).
	
	We also consider information pooling at the forecasting stage, by calculating forecast combinations. Forecast combinations find widespread application in the forecasting literature (\citealt{ElliottTimmermann2016}, \citealt{ChauvetPotter2013}) on the grounds that individual models are likely to be misspecified and that combining forecasts from different models should increase efficiency compared to individual forecasts. Measuring the performance of combined forecasts helps give a sense of how useful it is in practice to establish rankings of predictors: if combined forecasts turn out to outperform forecasts from every individual model, then the information contained in poorly-ranking predictors should not be discarded. 
	We combine the forecasts produced by single-predictor VAR models using two different weighting schemes. Let $w_i$ denote the weight assigned to model $i$ and $M$ the number of models to be combined.  
	The simplest approach consists in using an equal-weighted average of the forecasts, i.e., $w_i^{equal} = 1/M$.
	There is ample empirical evidence that equal weighting performs well for point forecasts (e.g., \citealt{StockWatson2003}) and often outperforms more sophisticated weighting strategies (\citealt{ElliottTimmermann2016}, \citealt{SmithWallis2009}).
	The second approach is Bayesian model averaging (BMA). In particular, since the value of the BIC for model $i$ provides an asymptotic approximation to its marginal likelihood, the BMA weights are approximated by
	$
	w_i^{bma} = \exp(-0.5 BIC_i) / [\sum_{i=j}^{M} \exp(-0.5 BIC_j) ]
	$.
	The weights are computed recursively across estimation windows.

	In addition, we report the performance of forecasts produced by a major forecasting institution, the International Monetary Fund (IMF).\footnote{The IMF forecasts are particularly suitable for comparisons in this context, as they cover horizons from 1 to 5 years. In the case of other major forecasters, comparisons would necessarily have limitations in terms of horizons. For instance, the Fed Greenbook forecasts cover a maximum horizon of 2 years. The OECD publishes annual forecasts for the following year and long-term projections in 10-year steps. The Survey of Professional Forecasters (SPF) includes quarterly forecasts up to 4 quarters ahead, annual forecasts for the next three years but only starting from the 2009Q2 survey, and 10-year annual average forecasts. However, as \citet{OECD2014} point out, ``the profile and magnitude of the errors in the GDP growth projections [over 2007-2012] of other international organizations and consensus forecasts are strikingly similar".}
	\footnote{The IMF World Economic Outlook (WEO) provides forecasts in the form of annual growth rates up to 5 years ahead. We convert them into 4-quarter-ahead, 12-quarter-ahead and 20-quarter-ahead forecasts of the GDP level using the following approach: (i) we take the Spring issues of the WEO; (ii) we consider the last quarter of the year prior to each issue as the starting value for forecasting; (iii) we apply the annual forecast growth rates to the starting value to compute forecasts of the GDP level and (iv) we assign each resulting value to the last quarter of the relevant forecast year. For example, the annual forecast growth rate for 1990 published in the 1990 Spring issue is used to compute the 4-quarter-ahead forecast for 1990Q4 based on historical data up to 1989Q4, the annual rates up to 1992 are used to compute the 12-quarter-ahead forecast for 1992Q4 based on data up to 1989Q4, and so on.}

	Table \ref{tab:mse_hd_models} reports the relative MSFE of all the alternative forecasting models/methods. No model nor forecast combination scheme outperforms the single-predictor models using CAPR and NNBLI at any forecast horizon. Also, unlike the small models using the best financial-cycle predictors, most models/combinations in Table \ref{tab:mse_hd_models} provide their best results relative to the benchmark AR at the shorter (4-quarter) horizon, with the exception of the factor model.

	Table \ref{tab:dm_hd_models} reports the p-values of the \citet{dieboldmariano1995} (DM) test of equal forecast accuracy,\footnote{We use the test correction proposed by \citet{HarveyLeybourneNewbold1997}.} making comparisons on a pairwise basis between the best-performing forecasts from Tables \ref{tab:mse_ardl}-\ref{tab:mse_var} and the forecasts by high-dimensional models/combinations considered in this section. Under the null hypothesis, competing forecasts provide equal MSFE, while under the alternative hypothesis forecasts by the best predictor have lower MSFE. At the 10\% significance level, the null is generally rejected for all horizons, except for forecast combinations and the LBVAR with $h=4$ (p-values are only slightly higher than 10\% in these cases).
	Overall, the DM test results corroborate our findings on the predictive importance of the CAPR and NNBLI ratios.
	In Section \ref{subsec:GR2010test}, we provide results from another test of forecast accuracy, the \citet{GiacominiRossi2010} test, taking into account forecast instabilities.

\begin{figure}[h]
		\caption{Forecasting the Great Recession: a comparison of models }
		\centering
		\includegraphics[scale=0.19]{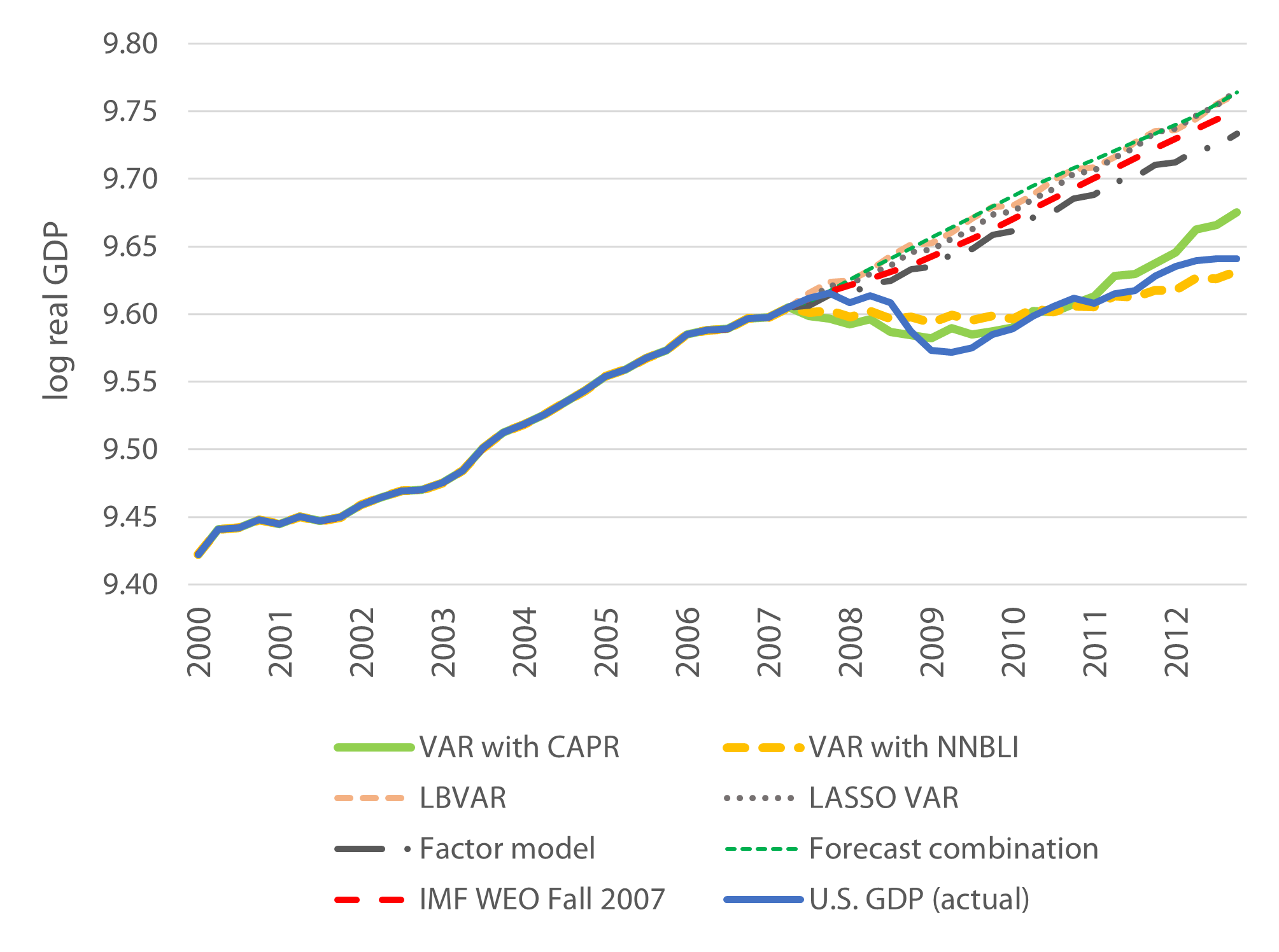}
		\subcaption*{\textit{Notes:} The figure shows the pseudo-out-of-sample forecasts of U.S. GDP over a 5-year horizon produced in 2007Q2 by alternative models: bivariate VAR with CAPR or NNBLI, Large Bayesian VAR (LBVAR), LASSO VAR, and a factor model. The figure also reports forecast combinations using equal weights and the forecasts contained in the Fall 2007 World Economic Outlook by the IMF. All models are estimated over the sample 1967Q1-2007Q2.}
		\label{fig:forecast_crisis}
	\end{figure}

	Figure \ref{fig:forecast_crisis} provides a focus on the Great Recession, to assess the ability of the models presented in this section to predict it. In particular, the figure compares the pseudo-out-of-sample iterated forecasts of GDP produced by alternative models in 2007Q2, with a horizon of 5 years. All models are estimated over the sample 1968Q2-2007Q2, i.e., using only observations prior to the beginning of the Global Financial Crisis.
	The figure also plots forecast combinations using equal weights and the forecasts published by the IMF in its Fall 2007 World Economic Outlook. 
	As shown in the figure, small models based on CAPR and NNBLI are much more effective in forecasting the Great Recession and the slow recovery than larger models.

	\begin{table}[H]
		\caption{High-dimensional models and forecast combinations: mean squared forecast error}
		\centering
		\scalebox{0.9}[0.9]{
		\begin{tabular}{c|ccc }
			\hline \hline 
			forecasting model/method	    & $h=4$      & $h=12$     & $h=20$     \\
			\hline
			LBVAR                                  & 0.82 & 0.90 & 0.75 \\
			LASSO VAR                              & 0.87 & 0.90 & 0.90 \\
			Factor model                           & 0.91 & 0.96 & 0.79 \\
			Forecast combination (equal   weights) & 0.93 & 0.99 & 1.00 \\
			Forecast combination (BMA   weights)   & 0.96 & 0.99 & 1.00 \\
			IMF                                    & 0.76 & 1.00 & 1.33 \\
			\hline \hline 
		\end{tabular}
		}
		\subcaption*{\textit{Notes:} The table shows the MSFE for the $h$-quarter-ahead cumulative GDP growth rate, relative to the benchmark AR model, over the period 1990Q1-2017Q4.
		The forecasting models/methods considered are: a Large Bayesian VAR (LBVAR); a VAR estimated using a LASSO penalty (LASSO VAR); a factor model; combinations of forecasts from one-predictor VAR models, using equal weights and Bayesian model averaging (BMA) weights; forecasts by the International Monetary Fund (IMF).
			}
		\label{tab:mse_hd_models}
	\end{table}

	\begin{table}[H]
		\caption{Diebold-Mariano test: best predictors vs. high-dimensional models}
		\centering
		\scalebox{0.9}[0.9]{
		\begin{tabular}{c|ccc}
			\hline \hline 
			forecasting model/method	   & $h=4$      & $h=12$     & $h=20$      \\
			\hline
			
			LBVAR                                & 0.12 & 0.07 & 0.04 \\
			LASSO VAR                            & 0.07 & 0.07 & 0.09 \\
			Factor model                         & 0.04 & 0.01 & 0.09 \\
			Forecast combination (equal weights) & 0.12 & 0.10 & 0.11 \\
			Forecast combination (BMA weights)   & 0.10 & 0.10 & 0.11 \\
			IMF                                  & 0.01 & 0.09 & 0.00\\
			\hline \hline 
		\end{tabular}
		}
		\subcaption*{\textit{Notes:} For each horizon $h$, the table reports the p-values of the Diebold-Mariano test of equal MSFE comparing the GDP forecasts using the best predictor from Table \ref{tab:mse_var} and those by high-dimensional models/methods from Table \ref{tab:mse_hd_models}. 
			Under the null hypothesis, competing forecasts have equal MSFE.
			Under the alternative hypothesis, forecasts by the best predictor have lower MSFE.
			All MSFE are computed over the period 1990Q1-2017Q4.}
		\label{tab:dm_hd_models}
	\end{table}

	\section{Extensions and robustness checks}\label{sec:robustness}
	
	In this section, we present a number of extensions to our main results as well as many robustness checks.
	
	First, we investigate whether the predictive power of CAPR and NNBLI is limited to recession periods, perhaps characterized by financial tightening.
	The recent literature on ``growth at risk" (\citealt{AdrianBoyarchenkoGiannone2019}; \citealt{AdrianGrinbergLiangMalikYu2022}) has found that the NFCI index, capturing general financial conditions, is useful to forecast tail risks to GDP, but is a weak predictor of the rest of the conditional GDP distribution.
	Following this literature, in Section \ref{subsec:quantile} we estimate quantile regressions for GDP growth, and show that the strong predictive relationship of CAPR and NNBLI with economic activity is stable across lower and upper percentiles of the GDP distribution (especially for CAPR), i.e., it appears to be present during both expansion and recession periods, unlike for NFCI. 
	
	Second, we assess variables' ability to specifically forecast the year-on-year growth rate at quarter $t+h$, instead of the growth rate over $h$ quarters (i.e., cumulative growth, and hence the $h$-quarter-ahead GDP level) considered so far, and show that the best forecasts are still provided by VAR models using CAPR and NNBLI.
	
	Next, we address the critical issue of forecast instabilities. A large literature has shown that variables' forecasting power is generally unstable over time, i.e., that some predictors provide good forecast performance, but only in specific periods (\citealt{Rossi2021}, \citealt{StockWatson2003}, \citealt{ClementsHendry2006}, among others).
	More specifically, we address the questions: To what extent are the main results determined by the Global Financial Crisis and Great Recession period? Would it have been possible to identify CAPR and NNBLI as top predictors in previous periods?
	Also, there exists a literature that focuses on parameter instabilities and finds that time-varying-parameter (TVP) models, in particular TVP-VARs, tend to produce better forecasts than models with time-invariant parameters (e.g., \citealt{KoopKorobilis2013}; \citealt{DAgostinoGambettiGiannone2013}).
	We assess the robustness of our main results to forecast and parameter instabilities in several ways.
	In Section \ref{subsec:rolling}, we repeat the out-of-sample evaluation of Section \ref{subsec:oos} estimating the models on rolling windows, which accommodate structural breaks in parameters.
	In Section \ref{subsec:tvp}, we consider forecasts produced by TVP-VARs.
	In Section \ref{subsec:subsamples}, we conduct in-sample and out-of-sample evaluation over sub-periods.
	In Section \ref{subsec:GR2010test}, we use the \citet{GiacominiRossi2010} test to account for instabilities when comparing forecast performance.
	
	Finally, in Section \ref{subsec:variable_selection}, we compare different variable-selection methods, to check if they select the same predictors, in particular CAPR and NNBLI.
	
	To facilitate reading, in all the tables reported in the following sections, we highlight CAPR and NNBLI using bold text.
	Also, in the interest of space, we place several tables and figures in the Online Appendix.

	\subsection{Quantile regressions}\label{subsec:quantile}
	
	First, we explore the predictive ability of CAPR and NNBLI at different quantiles of the conditional distribution of GDP growth. 
	To this aim, we first estimate quantile regressions and evaluate predictors using the local goodness-of-fit measure at the specific quantiles introduced by \citet{KoenkerMachado1999}.
	The estimated models are the same ARDL models as in Section \ref{eq:ardl}.
	Table \ref{tab:fit_qreg2} reports results for the 10th and 90th percentiles of the $h$-period GDP growth (very similar results are obtained for the 5th and 95th percentiles), listing only the top 5 predictors in each case to save space.
	CAPR is among the most powerful predictors for both quantiles, while
	NNBLI is a top predictor for the upper quantile but not also for the lower (although its results are not very far from the best ones for $h=20$ and $h=12$).
	
	Interestingly, for the lower quantile and for $h=4$, powerful predictors include a term spread ($T5YFFM$), which has long been used as a leading indicator of recessions (\citealt{StockWatson2003};  \citealt{ChauvetPotter2013}), and the composite National Financial Conditions Index (NFCI) (6th in the ranking, with a pseudo $R^2$ of 0.37, hence not reported in Table \ref{tab:fit_qreg2}), which are both weaker predictors of the upper quantile.
	This is consistent with the growth-at-risk literature (\citealt{AdrianBoyarchenkoGiannone2019}), which has previously shown the NFCI to provide valuable predictive content only for the left tail of the conditional GDP distribution. 
	\citet{ReichlinRiccoHasenzagl2019} further investigate this result by distinguishing between different components of NFCI. They find that price variables, such as credit spreads, actually provide limited information on growth vulnerability, whereas nonfinancial leverage (which is related to the NNBLI ratio) has provided useful early warnings for GDP in Global Financial Crisis. 
	
\begin{figure}[H]
		\caption{Coefficients of quantile regressions of GDP}
		\centering
		\begin{subfigure}[b]{0.32\textwidth}
			\centering
			\vspace{1pt}
			\caption*{NFCI, $h=4$}
			\vspace{-6pt}
			\includegraphics[width=\textwidth]{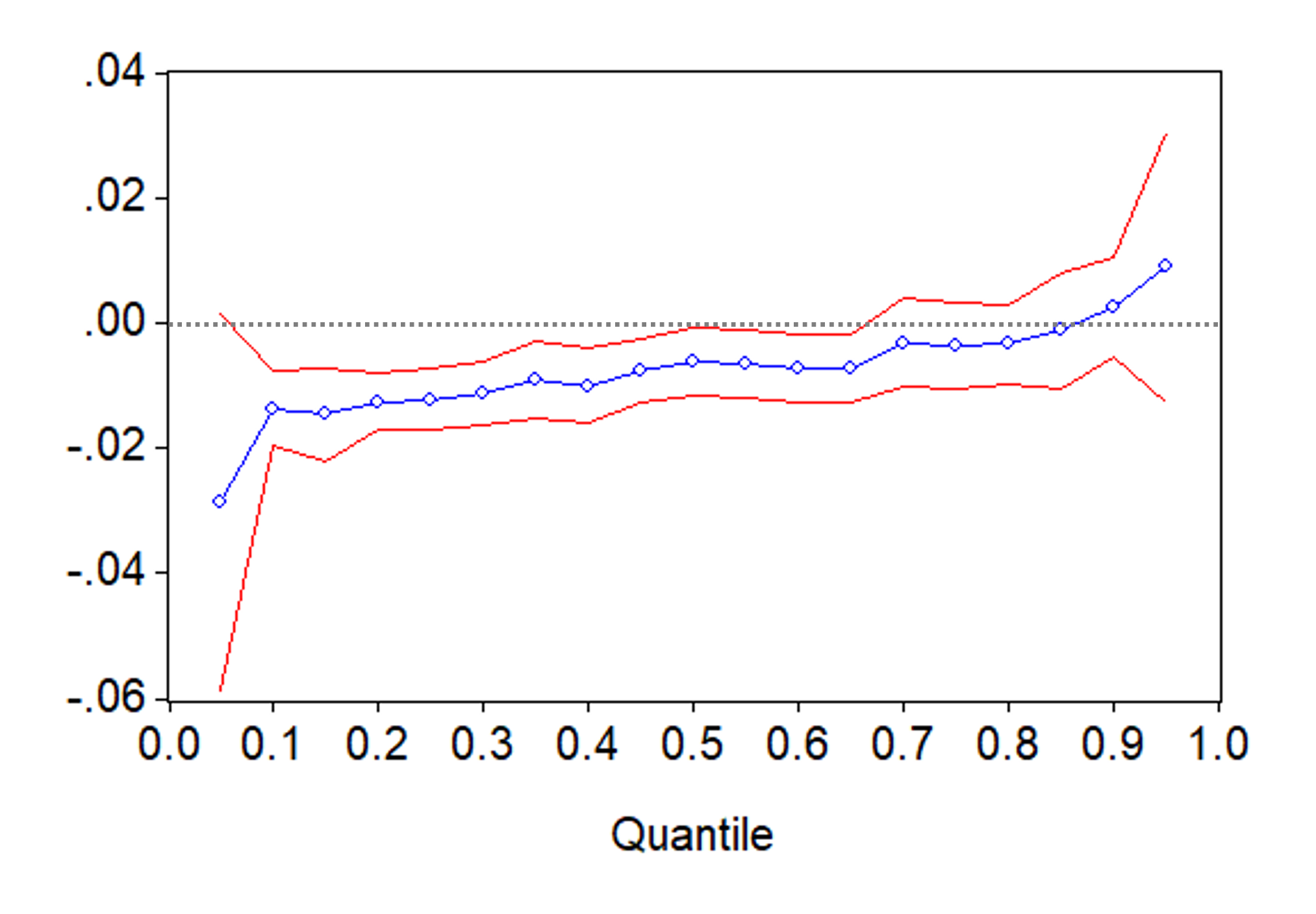}
		\end{subfigure}
		\begin{subfigure}[b]{0.32\textwidth}
			\centering
			\vspace{1pt}
			\caption*{NFCI, $h=12$}
			\vspace{-6pt}
			\includegraphics[width=\textwidth]{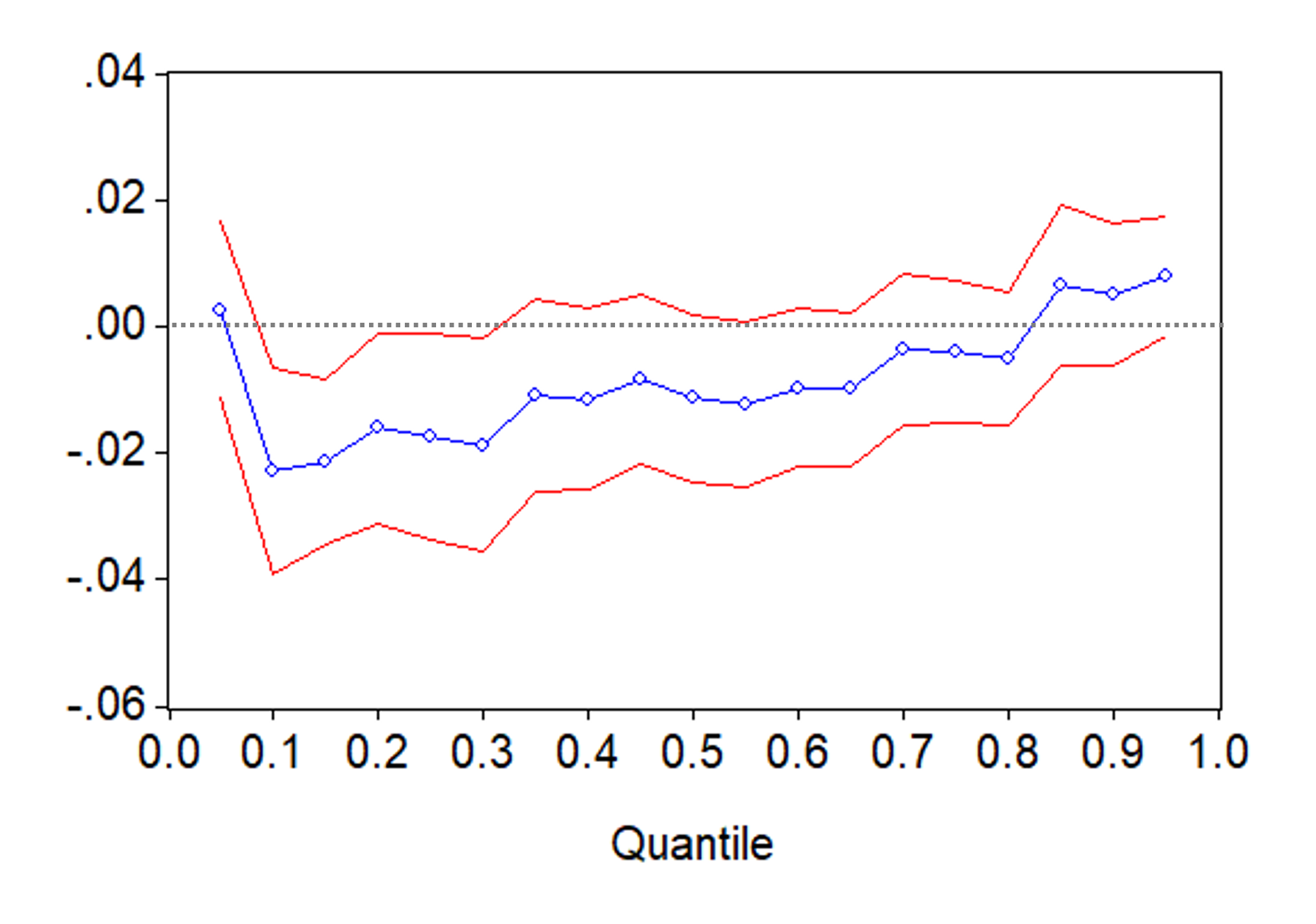}
		\end{subfigure}
		\begin{subfigure}[b]{0.32\textwidth}
			\centering
			\vspace{1pt}
			\caption*{NFCI, $h=20$}
			\vspace{-6pt}
			\includegraphics[width=\textwidth]{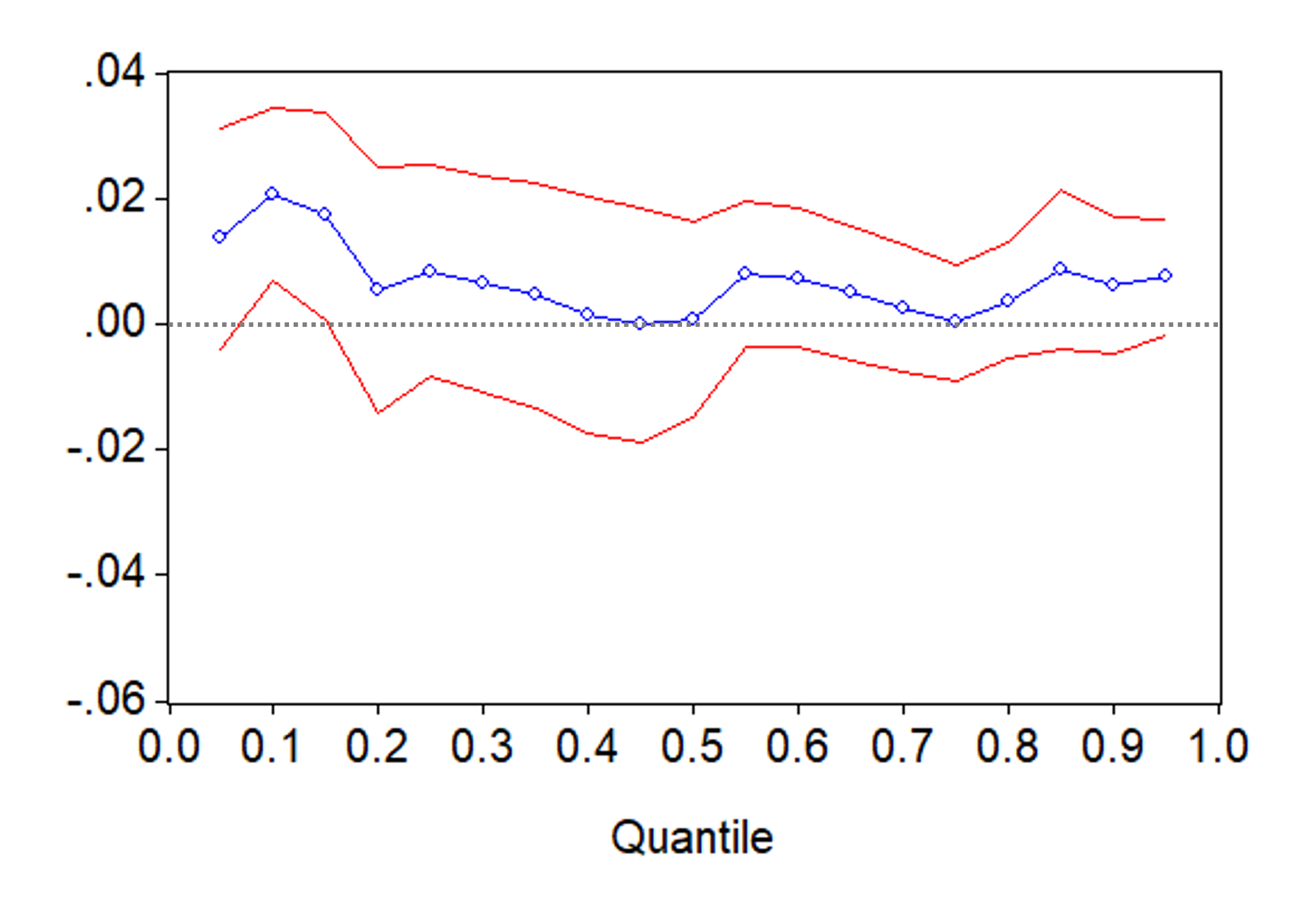}
		\end{subfigure}
		\begin{subfigure}[b]{0.32\textwidth}
			\vspace{1pt}
			\caption*{CAPR, $h=4$}
			\vspace{-6pt}
			\centering
			\includegraphics[width=\textwidth]{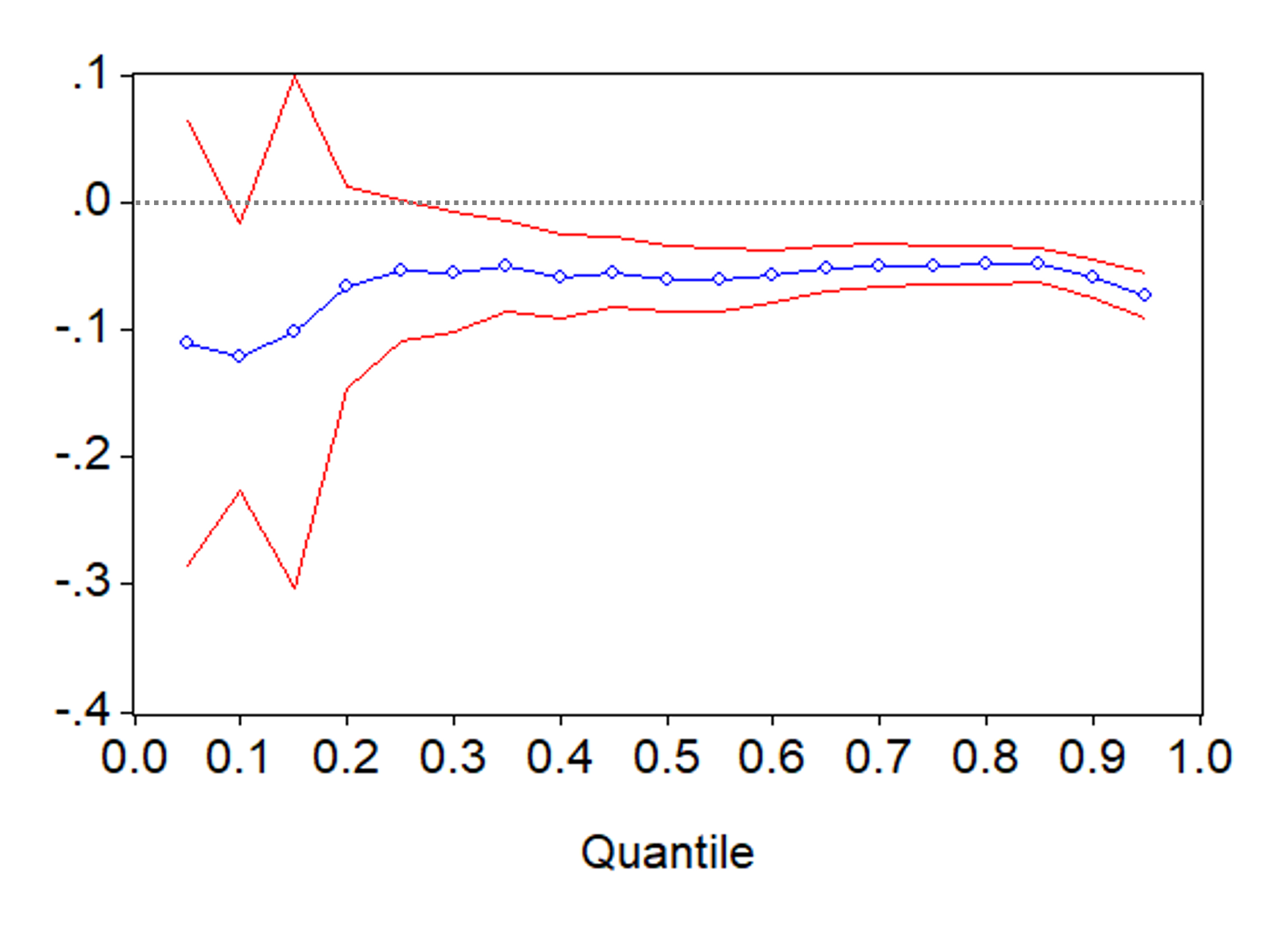}
		\end{subfigure}
		\begin{subfigure}[b]{0.32\textwidth}
			\vspace{1pt}
			\caption*{CAPR, $h=12$}
			\vspace{-6pt}
			\centering
			\includegraphics[width=\textwidth]{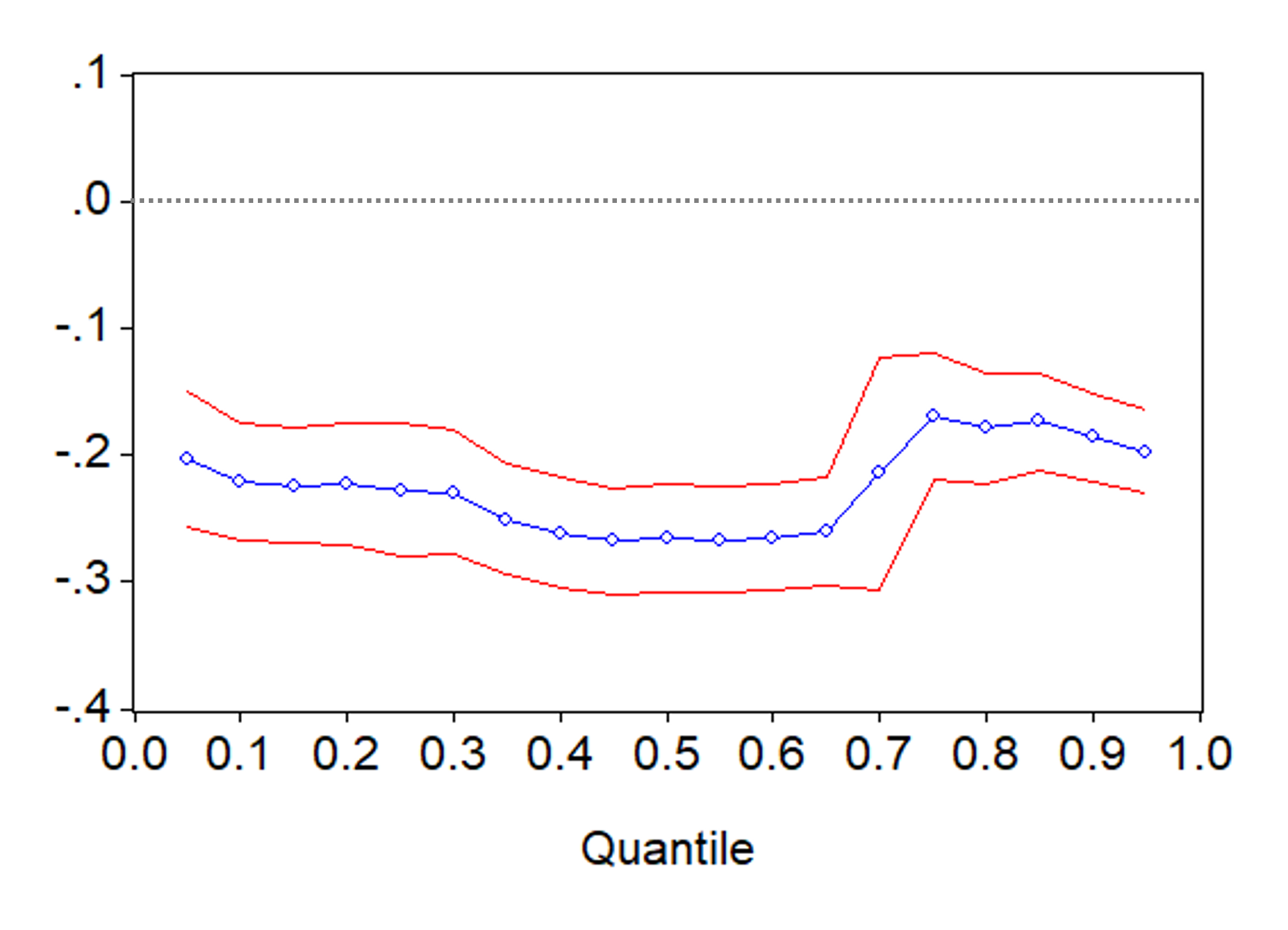}
		\end{subfigure}
		\begin{subfigure}[b]{0.32\textwidth}
			\vspace{1pt}
			\caption*{CAPR, $h=20$}
			\vspace{-6pt}
			\centering
			\includegraphics[width=\textwidth]{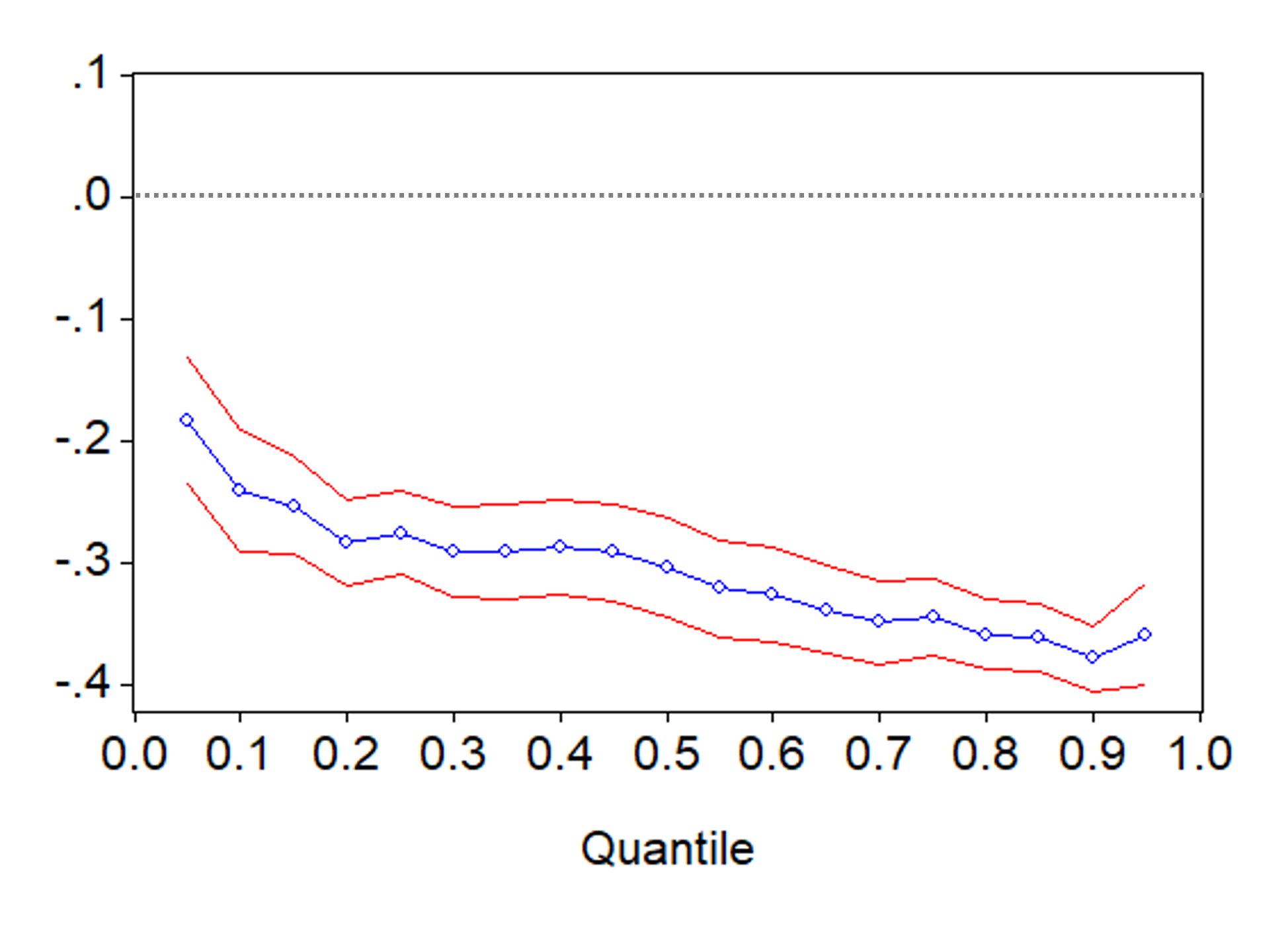}
		\end{subfigure}
		\begin{subfigure}[b]{0.32\textwidth}
			\vspace{1pt}
			\caption*{NNBLI, $h=4$}
			\vspace{-6pt}
			\centering
			\includegraphics[width=\textwidth]{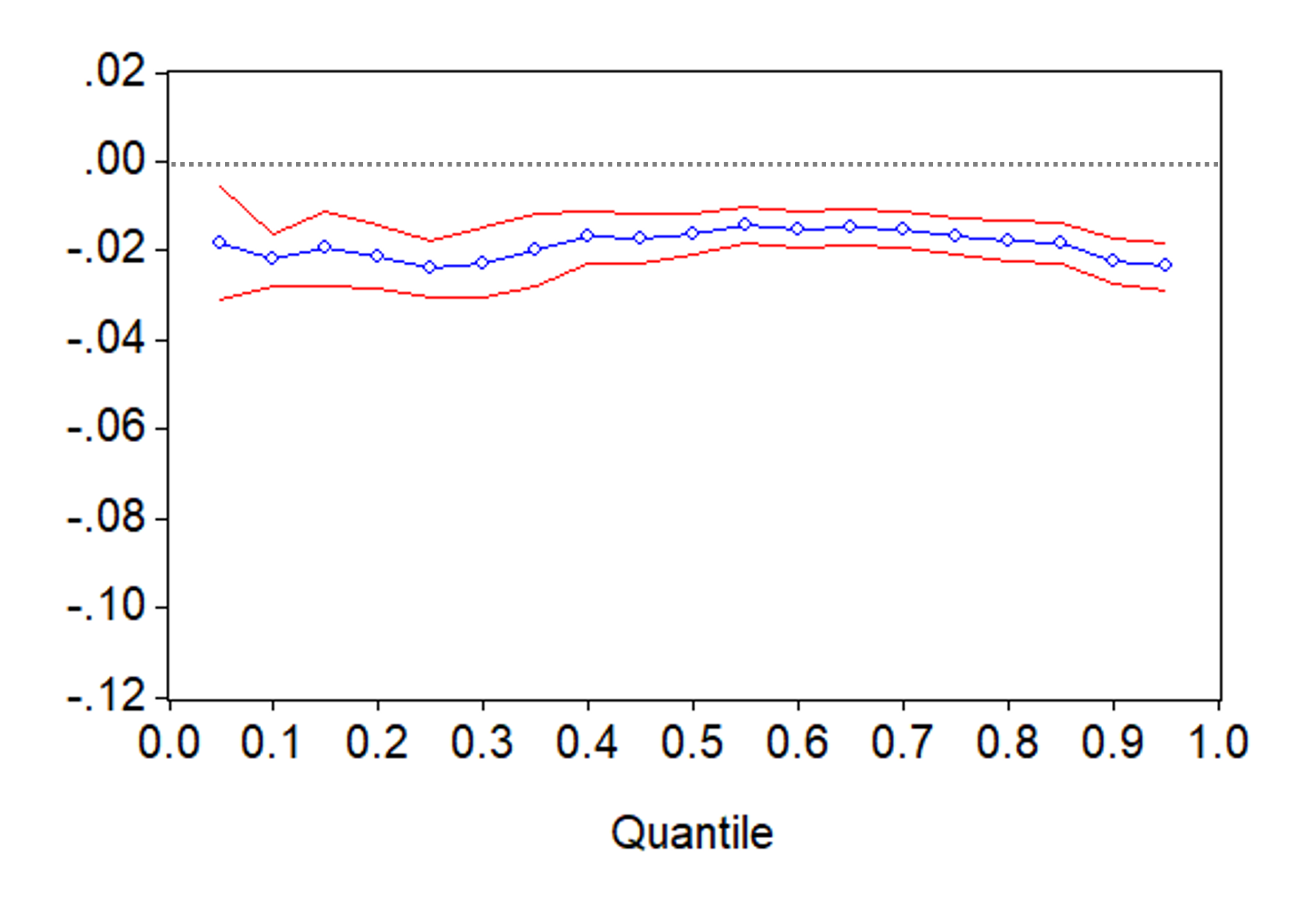}
		\end{subfigure}
		\begin{subfigure}[b]{0.32\textwidth}
			\vspace{1pt}
			\caption*{NNBLI, $h=12$}
			\vspace{-6pt}
			\centering
			\includegraphics[width=\textwidth]{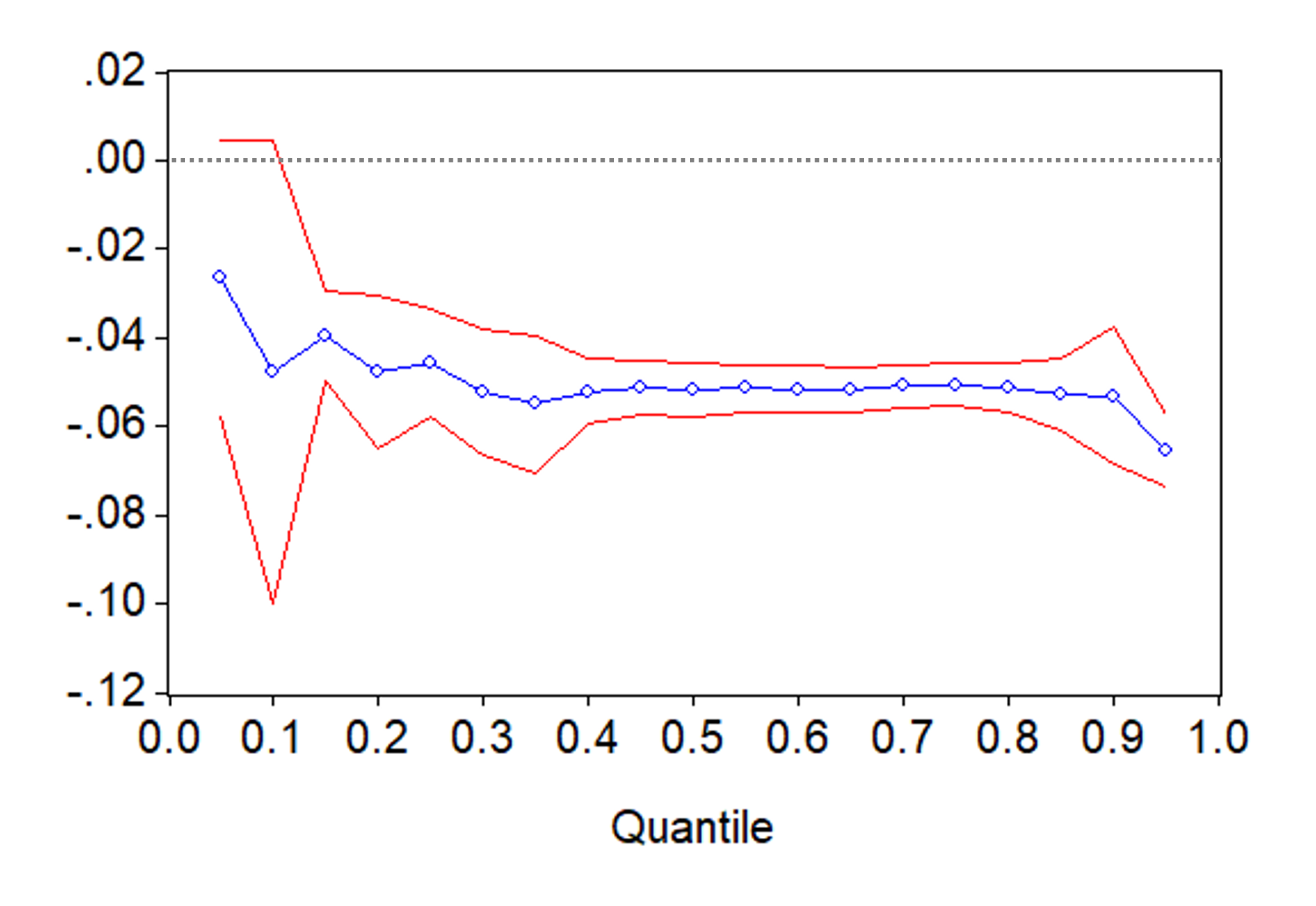}
		\end{subfigure}
		\begin{subfigure}[b]{0.32\textwidth}
			\vspace{1pt}
			\caption*{NNBLI, $h=20$}
			\vspace{-6pt}
			\centering
			\includegraphics[width=\textwidth]{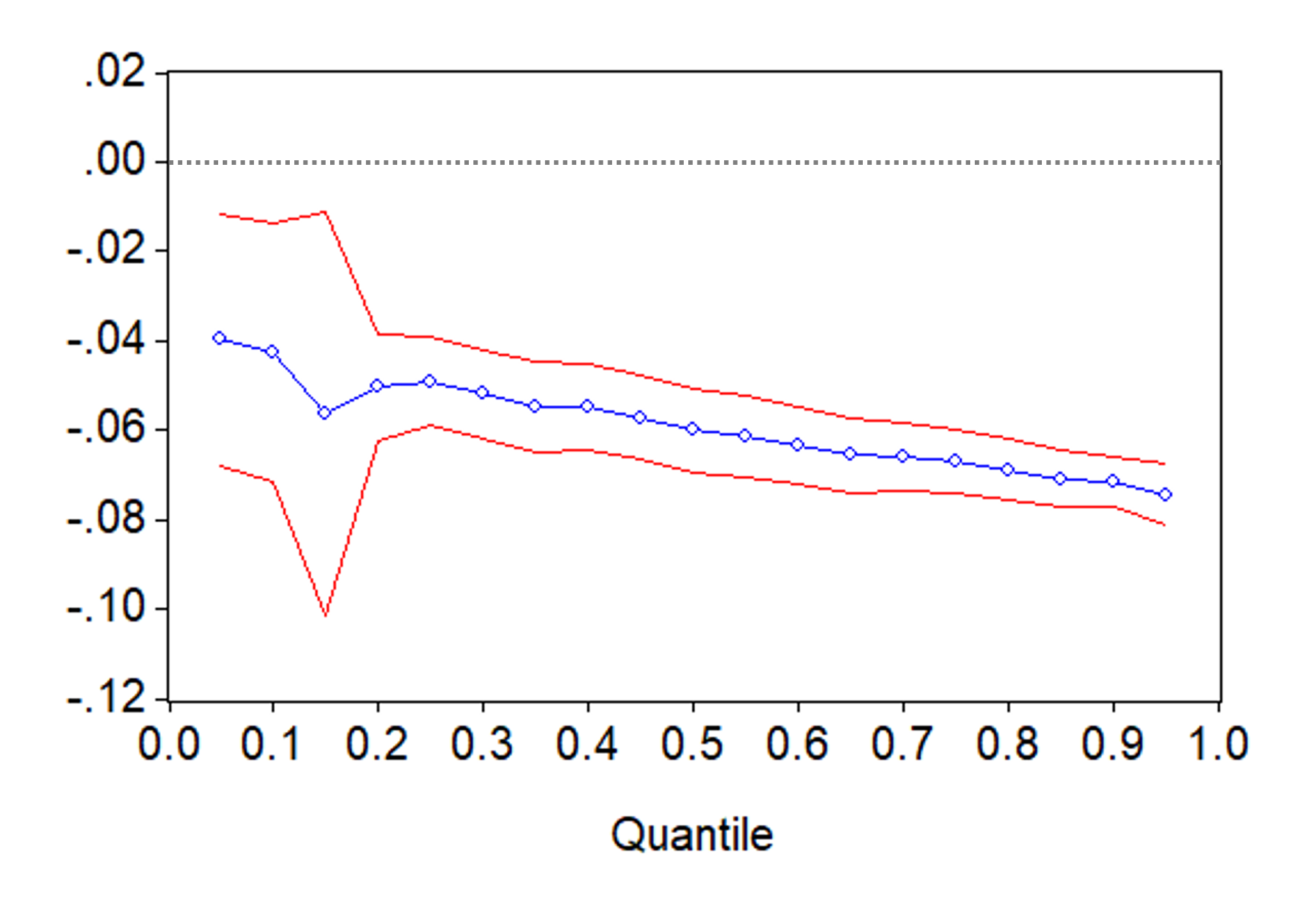}
		\end{subfigure}
		\label{fig:qreg_coeffs_2}
		\subcaption*{\textit{Notes:} The figure shows the estimated coefficients associated with different predictors in quantile regressions for $h$-quarter cumulative GDP growth (with 95\% confidence intevals).
		The variable NNBLI is divided by 100,000 to facilitate reading of the estimates.
		Dotted horizontal lines indicate the value of zero.}
	\end{figure}

	To better characterize our findings and to further connect to the literature on growth vulnerabilities, we next consider the coefficients of the quantile regressions of GDP using (alternatively) NFCI, CAPR and NNBLI as predictors, and we check if and how these coefficients vary across quantiles of GDP.
	Figure \ref{fig:qreg_coeffs_2} displays the coefficients for different quantiles of GDP from 5\% to 95\%.
	We report results for the simplest ARDL regressions with only one $h$-quarter lagged term for each predictor, i.e., equation \eqref{eq:ardl} with $p=q=1$.  
	For each regression, the figure reports the point estimate of the coefficient (blue line) and the 95\% confidence intervals (red lines). A dotted horizontal line indicates the value of zero.
	The results are quite revealing.
	NFCI has negative and significant coefficients only for lower percentiles of GDP over 1-3 year horizons (meaning that higher levels of financial stress predict lower GDP growth in the following periods), while coefficients are non-significant and positive in point estimate for higher percentiles and for all percentiles at the longest horizon $h=20$. These results are broadly consistent with those by \citet{AdrianBoyarchenkoGiannone2019} and \citet{AdrianGrinbergLiangMalikYu2022}.
	Conversely, the coefficients on CAPR and NNBLI are invariably negative across all quantiles and horizons, and in general strongly significant (with a few exceptions at low quantiles), their absolute magnitude increasing with $h$. 
	Thus, the relationship of CAPR and NNBLI with GDP appears stable across different parts of the GDP distribution. 
	Overall, these results indicate that the predictive power of the two ratios is not simply related to recessions induced by financial distress, unlike for measures of general financial conditions.
	The results presented in the remainder of the paper confirm that CAPR and NNBLI achieve good predictive performance during both recession and expansion periods.

	\begin{table}[H]
		\centering
		\caption{Goodness of fit of quantile regressions for GDP growth}%
		\scalebox{0.7}[0.7]{
			\begin{tabular}{l|ll|ll|ll}
				\hline \hline 
				& \multicolumn{2}{c|}{$h=4$}       & \multicolumn{2}{c|}{$h=12$} & \multicolumn{2}{c}{$h=20$}  \\
				\hline
				& \multicolumn{6}{c}{\textbf{Quantile: 10\%}} \\
				\hline
				1  & prfi_gdp   & 0.43 & AMDMUOx  		& 0.51 & pr              & 0.53 \\
				2  & PRFIx      & 0.42 & \textbf{capr}  & 0.45 & ISRATIOx        & 0.53 \\
				3  & T5YFFM     & 0.39 & pr       		& 0.43 & \textbf{capr}   & 0.51 \\
				4  & cli        & 0.38 & prfi_gdp 		& 0.43 & mortg_inc       & 0.48 \\
				5  & PERMITMW   & 0.37 & ISRATIOx 		& 0.42 & LIABPIx         & 0.48 \\
				\hline 
				& \multicolumn{6}{c}{\textbf{Quantile: 90\%}} \\
				\hline
				1  & cred_gdp        			& 0.38 & \textbf{capr}            	& 0.50 & \textbf{capr}            	& 0.51 \\
				2  & \textbf{capr}            	& 0.35 & \textbf{NNBTILQ027SBDIx} 	& 0.43 & \textbf{NNBTILQ027SBDIx} 	& 0.48 \\
				3  & \textbf{NNBTILQ027SBDIx} 	& 0.35 & pr              			& 0.41 & pr              			& 0.44 \\
				4  & LIABPIx         			& 0.35 & cli             			& 0.40 & cred_gdp        			& 0.40 \\
				5  & mortg_inc       			& 0.35 & cred_gdp        			& 0.40 & LIABPIx         			& 0.34 \\
				\hline \hline 
			\end{tabular}
		}
		\subcaption*{\textit{Notes:} The table reports the local goodness-of-fit measure (pseudo $R^2$) by \citet{KoenkerMachado1999} for quantile regressions.
			For each $h$, the dependent variable is GDP growth over $h$ quarters. All models are estimated using data from 1974Q1 to 2017Q4. Please refer to Tables \ref{tab:redux} and \citet{McCrackenNgFREDQD} for a description of the variables.}
		\label{tab:fit_qreg2}
	\end{table}

	\subsection{Forecasting year-on-year growth rate at horizon $h$}\label{subsec:yoy}
	
	Following the \citet{StockWatson2003} approach, this paper focuses on predicting GDP growth over $h$ periods (i.e., cumulative growth), and therefore the $h$-period-ahead GDP level.
	Multi-year growth also appears of special interest from more recent and related papers dealing with forecasts of GDP over medium-term horizons (e.g., \citealt{MianSufiVerner2017} on household debt and GDP, or \citealt{AdrianGrinbergLiangMalikYu2022} on the term structure of growth at risk). 
	As an extension to this main approach, we now evaluate forecasts of the specific growth rate of GDP at a given future quarter $t+h$, irrespective of GDP movements in previous periods.
	We report results on year-on-year growth rates, i.e., relative the same quarter of the previous year, but similar qualitative results are obtained for quarter-on-quarter growth rates.
	
	Table \ref{tab:mse_ardl_var_d4} reports the best ARDL and VAR forecasts for this target variable. 
	Of course, the results for $h=4$ coincide with those already presented in Tables \ref{tab:mse_ardl}-\ref{tab:mse_var}
	As Table \ref{tab:mse_ardl_var_d4} shows, the best forecasts are still provided by VAR models using CAPR and NNBLI.
	In particular, CAPR and the unadjusted price-rent ratio produce the best forecasts for both $h=12$ and $h=20$, while NNBLI is the best predictor at $h=4$ and the third-best performer for $h=12$. 
	ARDL forecasts do not achieve the same level of accuracy for this target variable.
	
	In the Online Appendix, we also report (Table \ref{tab:mse_hd_models_d4}) the relative MSFE of high-dimensional models/methods considered in Section \ref{subsubsec:hd_models}. 
	All models/methods perform better than the benchmark AR at short horizons, and LBVAR and LASSO VAR produce good forecasts at all horizons. Still, all models are outperformed by the best one-predictor models at all horizons.

	\begin{table}[H]
		\centering
		\caption{Forecasts of $h$-period-ahead year-on-year GDP growth rate: MSFE}%
		\scalebox{0.75}[0.75]{
			\begin{tabular}{l|ll|ll|ll}
				\hline \hline 
				& \multicolumn{2}{c|}{$h=4$}       & \multicolumn{2}{c|}{$h=12$} & \multicolumn{2}{c}{$h=20$}  \\
				\hline
				& \multicolumn{6}{c}{\textbf{ARDL}} \\
				\hline
				1  & \textbf{NNBTILQ027SBDIx} & 0.55 & AAA      & 0.93 & AAA      & 0.90 \\
				2  & \textbf{capr}            & 0.74 & cli      & 0.93 & BAA      & 0.92 \\
				3  & AMDMUOx         & 0.84 & gs10     & 0.93 & TCU      & 0.92 \\
				4  & pr              & 0.86 & GS5      & 0.94 & MZMREALx & 0.93 \\
				5  & cli             & 0.87 & BAA      & 0.95 & CUMFNS   & 0.94 \\
				\hline
				& \multicolumn{6}{c}{\textbf{VAR}} \\
				\hline
				1  & \textbf{NNBTILQ027SBDIx} 	& 0.55 & \textbf{capr}            	& 0.68 & \textbf{capr}   & 0.71 \\
				2  & \textbf{capr}            	& 0.71 & pr              			& 0.74 & pr              & 0.72 \\
				3  & pr              		  	& 0.76 & \textbf{NNBTILQ027SBDIx} 	& 0.88 & ISRATIOx        & 0.79 \\
				4  & AMDMUOx         		  	& 0.80 & ISRATIOx        			& 0.89 & B021RE1Q156NBEA & 0.89 \\
				5  & TLBSNNCBBDIx    			& 0.83 & UNRATESTx       			& 0.90 & REVOLSLx        & 0.93 \\
				\hline \hline 
			\end{tabular}
		}
		\subcaption*{\textit{Notes}: Out-of-sample MSFE for the $h$-quarter-ahead year-on-year GDP growth rate, relative to the benchmark AR. All models are estimated on recursive windows (shortest sample 1968Q2-1985Q1, longest sample 1968Q2-2016Q4) and MSFE are computed over the period 1990Q1-2017Q4. Please refer to Tables \ref{tab:redux} and \citet{McCrackenNgFREDQD} for a description of the variables.}
		\label{tab:mse_ardl_var_d4} 
	\end{table}

	\subsection{Forecasts using rolling-window estimates}\label{subsec:rolling}
	
	We now go back to our main goal of forecasting GDP growth over $h$ periods, and assess the extent to which the out-of-sample evaluation of Section \ref{subsec:oos} is robust to estimation on different samples. 
	
		In particular, we calculate the relative MSFE of forecasts produced by ARDL and VAR models using rolling windows with a fixed length instead of the recursive windows used for the baseline results of Section \ref{subsec:oos}. We consider rolling windows of size 40, 60 and 80 quarters, and report results for the mid size, 60 (in the Online Appendix, Table \ref{tab:mse_ardl_var_rolling}, which shows the top 10 predictors and the associated MSFE for each horizon).
	Thus, for instance, forecasts generated at time 2007Q2 are based on estimates obtained over the sample 1992Q3-2007Q2 instead of 1968Q2-2007Q2.
		The MSFE is calculated relative to the benchmark (recursive-scheme) MSFE of the AR model from Section \ref{subsec:oos}.
	As already mentioned, rolling windows provide less accurate top forecasts compared to recursive windows. \footnote{The accuracy of top forecasts increases with the size of the rolling windows. However, the results for size 80 quarters are similar to those for 60.}
	
	The results are in line with those of Section  \ref{subsec:oos}, especially for NNBLI. 
	Considering ARDL and VAR forecasts jointly, NNBLI is the best predictor for $h=4$ and $h=12$ and the second best for $h=20$. The unadjusted price-rent ratio is the best predictor for $h=20$, and CAPR is among the top predictors at all horizons.
	In the case of VAR forecasts,  NNBLI is the top performer at all horizons, while CAPR is second for $h=4$, 3rd for $h=20$, and 5th for $h=12$. 
	Other top predictors include the unfilled orders for
	durable goods (\textit{AMDMUOx}), nonfinancial corporate business sector net
	worth to disposable business income (\textit{TNWMVBSNNCBBDIx}), CLI and residential fixed investment (\textit{PRIx}).

	\subsection{Forecasting with time-varying parameter VARs}\label{subsec:tvp}
	
	The results obtained using rolling-window estimates suggest that our main findings are quite robust to parameter instabilities.
	In this section, we go a step further in this direction. 
	So far, we have considered models whose parameters do not vary within the estimation sample.
	As already mentioned, there exists a literature that finds time-varying-parameter VARs (TVP-VARs) to produce better forecasts than models with time-invariant parameters (e.g., \citealt{KoopKorobilis2013}; \citealt{DAgostinoGambettiGiannone2013}).
	We now consider forecasts from TVP VARs and check whether our main results are confirmed. We evaluate the TVP-VARs using the baseline recursive-window scheme.
	
	We first consider bivariate TVP-VARs with GDP growth and one predictor at a time.\footnote{We estimate TVP-VARs à la \citet{Primiceri2005}, which also allow for stochastic volatility of errors. The VAR coefficients are assumed to follow a random walk process, so their best estimate for out-of-sample periods corresponds to the value estimated for the last quarter of the sample window. We estimate models using code by Gary Koop, available at \url{https://sites.google.com/site/garykoop/}, and report results based on 1000 Markov Chain Monte Carlo (MCMC) replications and 1000 burn-in replications. To reduce the computational burden, we do not implement automatic lag selection as in Section \ref{subsec:oos}, but assume a fixed lag order of 2 in all models. In each sample window, the prior for the TVP-VAR is given by the OLS estimates of time-invariant parameters obtained using data within that window. We also consider uninformative priors, which however produce less accurate top forecasts.}  
	The upper part (Panel A) of Table \ref{tab:mse_tvpvar} reports the results for the top 5 predictors. The MSFE are expressed relative to the benchmark time-invariant-parameter AR model.
	The table broadly confirms our main findings: CAPR and NNBLI are the best two predictors at all horizons. 
	Using TVP-VARs further improves the forecast performance of NNBLI for $h=4$ and $h=12$, compared to Table \ref{tab:mse_var}, while forecasts of both CAPR and NNBLI for $h=20$ are worse than their time-invariant-parameter counterparts.

	Next, we extend our set of high-dimensional models by considering the Large TVP-VAR approach by \citet{KoopKorobilis2013}.
	In particular, these authors propose dynamic model selection (DMS) and dynamic model averaging (DMA) methodologies to mix predictions from TVP models of different sizes. 
	We implement the TVP-VAR-DMS and TVP-VAR-DMA approaches using two TVP-VARs.\footnote{We use software code provided by \citet{KoopKorobilis2013}. Based on their results, we set the ``forgetting factors" to the value of 0.99 and the ``decay factor" to 0.96 (please see \citealt{KoopKorobilis2013} for details).} The smaller TVP-VAR is the 3-variable model used by \citet{KoopKorobilis2013}, which includes GDP, inflation and interest rate. The larger TVP-VAR includes the set of 18 predictors selected by the LASSO estimator (over the full sample) in the GDP equation of the LASSO VAR of section \ref{subsubsec:hd_models} (the list of predictors is presented in Section \ref{subsec:variable_selection}). 
	We consider lag orders from 1 to 5 and only report the best results for each forecast horizon.
	The lower part (Panel B) of Table \ref{tab:mse_tvpvar}
	displays the MSFE of TVP-VAR-DMA, which provides better forecasts than TVP-VAR-DMS in this context. 
	The approach achieves good performance for $h=4$, with a relative MSFE of around 0.9, close to the values obtained by high-dimensional models from Table \ref{tab:mse_hd_models} (in particular, LBVAR and factor model), whereas its forecast accuracy deteriorates at longer horizons.
	More importantly, it is outperformed by the best one-predictor VARs (both time-invariant and time-varying) at all horizons.
	This is in part consistent with \citet{KoopKorobilis2013}, who find that small TVP-VARs tend to be preferred to larger models when forecasting GDP (while larger TVP-VARs and TVP-VAR-DMS/DMA are more useful for inflation and interest rates).

	\begin{table}[H]
		\centering
		\caption{TVP-VAR forecasts: MSFE}%
		\scalebox{.85}[.85]{
			\begin{tabular}{l|ll|ll|ll}
				\hline \hline 
				& \multicolumn{2}{c|}{$h=4$}       & \multicolumn{2}{c|}{$h=12$} & \multicolumn{2}{c}{$h=20$}   \\
				\hline 
				& \multicolumn{6}{c}{\textit{Panel A:} \textbf{Bivariate TVP-VAR}} \\
				\hline
				1 & \textbf{NNBTILQ027SBDIx} & 0.53 & \textbf{NNBTILQ027SBDIx} & 0.28 & \textbf{capr}           & 0.46 \\
				2 & \textbf{capr}           & 0.71 & \textbf{capr}           & 0.57 & \textbf{NNBTILQ027SBDIx} & 0.50 \\
				3 & \textbf{pr}         	& 0.79 & \textbf{pr}         	& 0.73 & \textbf{pr}         & 0.57 \\
				4 & prfi\_gdp       		& 0.80 & AMDMUOx         		& 0.73 & NWPIx           & 0.78 \\
				5 & hpi		   	    		& 0.81 & cli             		& 0.79 & AMDMUOx         & 0.80 \\
				\hline \hline
				& \multicolumn{6}{c}{\textit{Panel B:} \textbf{Large TVP-VAR}} \\
				\hline
				&  \multicolumn{2}{c}{0.91}  & \multicolumn{2}{c}{0.98}  & \multicolumn{2}{c}{1.14} \\
				\hline \hline 
			\end{tabular}
		}
		\subcaption*{\textit{Notes}: 
			The table reports the MSFE for the $h$-quarter (cumulative) GDP growth rate, relative to the benchmark AR, computed over the period 1990Q1-2017Q4, 
			using bivariate TVP-VAR models (Panel A) 
			and the TVP-VAR-DMA approach by \citet{KoopKorobilis2013} (Panel B). 
			 See Table \ref{tab:redux} and \citet{McCrackenNgFREDQD} for a description of the predictors listed in Panel A. See section \ref{subsec:tvp} for details on the TVP-VAR-DMA approach.}
		\label{tab:mse_tvpvar} 
	\end{table}

	\subsection{Evaluation over sub-samples}\label{subsec:subsamples}
	
	As a third check on the robustness of results to instabilities in predictive power, we evaluate predictions over sub-periods. 
	In particular, we further address the questions: do the main results from Sections \ref{subsec:is} and \ref{subsec:oos} simply depend on the inclusion of the Global Financial Crisis (GFC) and Great Recession in our sample? Would it have been possible to identify CAPR and NNBLI as top predictors before the GFC?
	
	To begin with, we calculate the $R^2$ of ARDL models from Section \ref{subsec:is} excluding the period 2007Q3-2009Q2, i.e., the GFC and Great Recession period.
	We take the third quarter of 2007 as the beginning of the GFC (in particular, the month of August, when BNP Paribas stopped withdrawals from three of its hedge funds, and major indicators of financial stress rose, e.g., the NFCI). The second quarter of 2009 is the end of Great Recession,  according to the NBER recession dates.
	The results are reported in the upper part of Table \ref{tab:r2_subsamples} in the Online Appendix. CAPR, the unadjusted price-rent (PR) ratio and NBBLI are the top three predictors for $h=12$ and $h=20$, and CAPR is also still among the top predictors for $h=4$.
	Next, we calculate the $R^2$ on periods before the GFC. 
	Part (b) of Table \ref{tab:r2_subsamples} in the Online Appendix shows the results obtained using data from 1974Q1 to 2007Q2 as in Section \ref{subsec:is}, while part (c) shows the results over a sample corresponding (approximately) to the ``Great Moderation" period, 1983Q1-2007Q2.
	CAPR is invariably among the top predictors for $h=12$ and $h=20$, while NNBLI provides the best fit for $h=12$ and the second-best fit for $h=4$ during the Great Moderation period. 	
	
	We then consider pseudo-out-of-sample forecasts by ARDL and VAR models, and split the forecast evaluation period in a pre-crisis period (1990Q1-2007Q2), a crisis period (2007Q3-2009Q2), and a post-crisis period (2009Q3-2017Q4).
	Tables \ref{tab:mse_ardl_subsamples}-\ref{tab:mse_var_subsamples} in the Online Appendix report the MSFE over these different periods. Overall, the forecasting power of CAPR and NNBLI appears stable compared to the other predictors.
	In the pre-crisis period, the best forecasts are provided by NNBLI at all horizons (in particular, VAR forecasts for $h=4$ and ARDL forecasts for $h=12$ and $h=20$), while
	CAPR appears especially useful for long horizons: it is the second-best predictor for $h=20$ (using ARDL forecasts), and among the best predictors for $h=12$.
	During the crisis period, the best forecasts for $h=4$ are provided by CAPR and NNBLI, along with private residential fixed investment (\textit{PRFIx}). For $h=12$, the best forecasts are again provided by CAPR and NNBLI, along with house price growth and the ratio of private residential fixed investment to GDP. CAPR also provides the best forecasts for $h=20$ (ARDL).
	In the post-crisis period, the two ratios are again the best predictors (both ARDL and VAR) for $h=12$ and $h=20$, together with the unadjusted price-rent ratio, while they are relatively less powerful at the shorter horizon $h=4$.\footnote{When the MSFE are calculated over a sample that only excludes the crisis period (not reported in the interest of space), CAPR is still the best predictor in the case of both ARDL and VAR forecasts for $h=20$, NNBLI is the best predictor and CAPR the second-best for $h=12$, and NNBLI is the best predictor (using both ARDL and VAR) for $h=4$.} 
	
	Overall, NNBLI-based forecasts appear better for horizons of 1 to 3 years, CAPR-based ones for 3 to 5 years. NNBLI would be unambiguously selected before the GFC for all horizons, while CAPR only in the case of the longest horizon.
	
	Finally, we focus on the ability of the two ratios to forecast recessions.
	We already showed in Figure \ref{fig:forecast_crisis} that both CAPR and NNBLI were remarkably effective in forecasting the Great Recession. In the Online Appendix (Figure \ref{fig:forecast_recessions}), we assess their ability to also forecast other recessions, namely the 1990-1991 recession and the 2001 recession. In particular, we consider forecasts produced by bivariate VAR models using either CAPR or NNBLI, estimated on data up to 1990Q2 or up to 2000Q4.
	In both cases, the two variables predict slowdowns in economic activity.
	NNBLI is more effective than CAPR in forecasting the 2001 recession.

	\subsection{Testing forecast accuracy in the presence of instabilities: the Giacomini-Rossi (2010) test}\label{subsec:GR2010test}
	
	\citet{GiacominiRossi2010} have proposed a methodology to test competing models' forecasting performance in a way that is robust to the presence of instabilities.
	As a further check, for each forecast horizon $h$, we take the best forecasts from Tables \ref{tab:mse_ardl}-\ref{tab:mse_var} (i.e., NNBLI-based VAR forecasts for $h=4$ and $h=12$ and CAPR-based ARDL forecasts for $h=20$) and perform the \citet{GiacominiRossi2010} test comparing these one-predictor forecasts with the forecasts produced by the high-dimensional approaches considered in Section \ref{subsubsec:hd_models}.
	
	Table \ref{tab:gr2010test} reports the values of the test statistics, along with 5\% and 10\% critical values.\footnote{The Giacomini-Rossi test is based on forecast loss differences between competing forecasts, calculated over rolling windows of out-of-sample observations. In our case, the time series of forecast errors run from 1990Q1 to 2017Q4 (112 quarters). We calculate loss (squared-error) differences on rolling widows of 40 quarters (so the first window is 1990Q1-1999Q4, the second is 1990Q2-2000Q1, and so on), but similar results are obtained using windows of 60 quarters. We calculate the test statistics using Newey-West heteroskedasticity and autocorrelation consistent (HAC) estimators of the long-run variances of the loss differences, with a bandwidth of 2 lags, based on the results from AR regressions, indicating that 2 lags generally capture all the significant autocorrelation in loss differences.} 
	The Giacomini-Rossi test results are broadly in line with the Diebold-Mariano test results from Table \ref{tab:dm_hd_models}.
	In particular, one-predictor model forecasts for $h=12$ and $h=20$ significantly outperform all forecasts from high-dimensional models at the 5\% level of significance and the IMF forecasts at the 10\% level, while forecast improvements are typically not statistically significant at the shorter horizon $h=4$, except relative to the IMF forecasts at the 10\% level.
	
	In the Online Appendix (Figure \ref{fig:gr2010test}), we provide a plot of the entire rolling sequence of out-of-sample loss differences between competing forecasts, along with the 5\% critical values of the test, from 1999Q4 to 2017Q4. 
	NNBLI-based forecasts for $h=12$ and CAPR-based forecasts for $h=20$ tend to significantly outperform forecasts by high-dimensional models/methods both before and after the Global Financial Crisis and Great Recession period.

	\begin{table}[H]
		\caption{Comparing forecast accuracy with instabilities: the Giacomini-Rossi (2010) test}
		\centering
		\scalebox{0.9}[0.9]{
			\begin{tabular}{cccccc|cc}
				\hline \hline 
				&      LBVAR    & LASSO VAR & Factor     & Combin. & IMF & 10\% c.v. & 5\% c.v.     \\
				\hline
				NNBLI, $h=4$ &2.424 & 2.126 & 2.364 & 2.058 & 2.630 & 2.626 & 2.890 \\
				NNBLI,  $h=12$   & 	3.185 & 5.518 & 6.639 & 4.138 & 2.830 & 2.626 & 2.890 \\
				CAPR, $h=20$ 	&6.404 & 4.346 & 3.511 & 4.661 & 2.843 & 2.626 & 2.890\\
				\hline 		\hline
			\end{tabular}
		}
		\subcaption*{\textit{Notes:} The table reports the Giacomini-Rossi (2010) test statistics, along with 5\% and 10\% critical values. The test is calculated using forecast errors in the period 1990Q1-2017Q4. Loss differences between competing models are calculated on rolling windows of 40 quarters. Under the null hypothesis, competing models have equal predictive ability.}
		\label{tab:gr2010test}
	\end{table}

	\subsection{Comparing variable-selection methods}\label{subsec:variable_selection}

	In section \ref{subsec:is}, predictors were considered one at a time, and CAPR and NNBLI stood out in the evaluation based on $R^2$.
	As a further robustness check, we now consider widely-used variable-selection methods, i.e., approaches for selecting subsets of predictors after pooling all the available data, to check whether they also select the same two predictors.\footnote{The pool of variables includes all FRED-QD variables, transformed as indicated by \citet{McCrackenNgFREDQD}, and the additional variables from Table \ref{tab:redux}.}
	We consider three different approaches: LASSO-based variable selection (\citealt{Tibshirani1996}), Bayesian variable selection using a popular shrinkage prior, namely the ``horseshoe prior" by \citet{CarvalhoPolsonScott2010}, and variable selection based on the random forest methodology (\citealt{GenuerPoggiTuleauMalot2010}), which is popular in machine learning.
	We apply these methods to ARDL models of GDP growth, considering all predictors at the same time. In the case of the LASSO, we also report the list of variables selected for the GDP equation of the LASSO VAR introduced in Section \ref{subsec:oos}, estimated on the largest sample.\footnote{
		For LASSO ARDL models, we first consider tuning the penalty parameter through a 10-fold cross-validation procedure, i.e., finding the value that minimizes the cross-validated MSE. However, this results in a large set of selected variables for all $h$, so for the ease of exposition -- and to more clearly highlight the importance of CAPR and NNBLI --, we report results using a stricter penalty ($\lambda = 0.005$), i.e., further shrinking the list of selected predictors.
		For LASSO VAR, we use the same value of the penalty parameter as in Section \ref{subsec:oos}, i.e., the one that provides the best forecasts of GDP.
		In the case of the horseshoe prior, variables are selected when the 90\% credible interval for their coefficient does not include zero.
		In more detail, the implemented horseshoe prior is based on a Cauchy prior truncated to [1/$k$, 1], where $k$ denotes the total number of predictors, and a Jeffreys prior for the error variance. The results are obtained using 5000 Markov Chain Monte Carlo (MCMC) samples and 2000 burn-in samples.
	}

	LASSO applied to ARDL selects NNBLI for all values of $h$ and CAPR for $h=12$ and $h=20$.
	The horseshoe prior selects NNBLI for all horizons, but not CAPR. 
	In the case of random forests, predictors are actually ranked (not just selected) on the basis of a measure of ``variable importance", capturing their contribution to the goodness of fit.\footnote{Specifically, the importance of each variable is measured as the increase in the ``out-of-bag" MSE resulting from a random permutation of that variable, averaged over all trees in the random forest. Only the best variables are kept, based on the thresholding strategy proposed by \citet{GenuerPoggiTuleauMalot2010}. Results are based on 5000 trees. 
	}
	This approach selects CAPR for all horizons, indicating it as the best predictor for both $h=12$ and $h=20$, while the unadjusted price-rent ratio is the best predictor for $h=4$. NNBLI is selected for $h=4$ and $h=12$.
	Finally, the LASSO VAR selects NNBLI but not CAPR.\footnote{It should be noted, however, that since VAR estimation by construction minimizes the 1-step-ahead mean squared error, the LASSO VAR approach may simply fail to recognize CAPR as a top predictor because this variable exhibits strong predictive power especially at long horizons. Moreover, each of the variables in the GDP equation has in turn its own equation in the LASSO VAR, so many more variables are indirectly involved in GDP forecasts in this case, as shown by Figure \ref{fig:lassovar_selection} introduced before.}
	Table \ref{tab:variable_selection} in the Online Appendix reports the complete list of variables selected by the different methods.

	Overall, these results confirm the importance of NNBLI and CAPR. 
	In particular, NNBLI is the only variable that is selected by all methods considered here, and for all horizons (except $h=20$ in the random forest). 
	As before, CAPR appears to be especially useful for horizons of 3 to 5 years.

	\subsection{Forecast evaluation using real-time data}\label{subsec:realtime}
	
	Lastly, we evaluate forecasts using historical data vintages, to check that our main results are not simply determined by the use of revised data.
	A fully real-time analysis is not feasible, due to limited availability of data vintages.
	In particular, the only vintage available for the FRED-QD dataset before 2018 is the beta version of November 2015.
	However, a large subset of variables is also included in the smaller FRED-MD dataset (\citealt{McCrackenNg2016}), for which a complete series of vintages is provided starting from September 1999.
	Therefore, we produce and evaluate real-time forecasts based on the complete set of vintages of FRED-MD (128 variables), using the 1999Q4 vintage for earlier periods.\footnote{In each quarter, we use the vintage released in the last month of the quarter. We use 1999Q4 as the first vintage, instead of 1999Q3, because many variables are not available in the 1999Q3 vintage.}
	We also check that our main results are confirmed when using the 2015-11 vintage of the complete FRED-QD dataset.

	For GDP, which is used both as the target variable and as a regressor in ARDL and VAR models, we retrieve the complete set of real-time vintages starting from December 1991 from the St. Louis Fed's Archival FRED (ALFRED) database. 
	We use real-time GDP data to produce forecasts and final (revised) data to evaluate them.

	For CAPR, \citet{DavisLehnertMartin2008} data on prices and rents for the aggregate stock of owner-occupied housing are not available in real time. However, similar values of the ratio are obtained using the Case-Shiller house price index and the Owners' Equivalent Rent of Residences in U.S. City Average by the U.S. Bureau of Labor Statistics, for which historical vintages are available in ALFRED, starting from April 2011 for rents and from November 2014 for house prices. 
	Moreover, for these series data revisions are generally negligible, so that CAPR data can be considered approximately real-time even before 2011 (to get real values, we divide by the non-seasonally-adjusted Consumer Price Index, which is not revised over time).\footnote{Since the FRED rent time series starts in 1983, to calculate CAPR before 1983 we extrapolate it backward using the growth rates of the data by \citet{DavisLehnertMartin2008}.}
	For NNBLI, we can only construct real-time vintages from 2010Q1, using raw data on noncorporate liabilities and business disposable income available in ALFRED (see \citealt{McCrackenNgFREDQD} for details on data elaboration).
	
	For the additional variables in Table \ref{tab:redux}, we collect real-time data from ALFRED when available. In general, the earliest vintages in ALFRED are only available after 2010.
	For the Composite Leading Indicator, real-time vintages are provided by the OECD, starting in May 2001.

	Table \ref{tab:mse_ardl_var_realtime} in the Online Appendix reports the results for ARDL and VAR forecasts using the available data vintages. 
	CAPR and NNBLI are confirmed as the top predictors at all horizons, along with the unadjusted price-rent ratio. 
	Although these results must be taken with caution, especially for NNBLI, they still suggest that the predictive power of the two ratios is quite robust to the use of alternative historical vintages.

	\section{Insights from macro-finance theory}\label{sec:theory}

	The results in the previous sections reveal that the CAPR and NNBLI ratios have a robust negative relationship with economic growth over the subsequent years.
	But theoretically, what justifies this predictive relationship? 
	Intuitively, since the two ratios combine information on financial conditions and economic fundamentals, they appear useful to signal financial vulnerabilities. 
	However, a discussion of the macro-finance theory provides deeper insights on the mechanisms linking these ratios to aggregate activity.

	The role of firms' debt as a key driver of business cycle fluctuations has long been recognized by macroeconomic theory (e.g., \citealt{BernankeGertlerGilchrist1999}; \citealt{KiyotakiMoore1997}).
	In particular, the ``financial accelerator" model by \citet{BernankeGertlerGilchrist1999}, arguably the most influential framework for studying the macro effects of business debt (\citealt{BrunnermeierKrishnamurthy2020}), hinges on the relationship between debt and firms' net worth, which is determined by business income and asset values. In the presence of credit market frictions, such as agency costs, lower (higher) net worth relative to debt leads to higher (lower) costs of financing, thereby decreasing (increasing) borrowing, investment and production. 
	As highlighted by \citet{BrunnermeierKrishnamurthy2020}, the financial accelerator framework is built on a corporate finance model (namely, an entrepreneur-manager firm model) that is most suitable for small firms. Since the standard organizational models of small firms are noncorporate ones, such as sole proprietorships and partnerships, this theoretical framework provides a strong rationale for the informative role of the debt-to-income ratio of noncorporate businesses, i.e., NNBLI.
	Moreover, in models like \citet{BernankeGertlerGilchrist1999} and \citealt{KiyotakiMoore1997}, durable goods, such as houses, are used as collateral for lending, so fluctuations in their prices represent an important amplification mechanism, affecting the availability of credit and production.
	Since collateralized borrowing is especially important for small businesses (e.g., \citealt{BanerjeeBlickle2021}), this framework also provides a straightforward theoretical link between the predictive power of NNBLI and that of housing market valuation metrics.\footnote{In the United States, mortgages represent a much larger share of debt for the noncorporate than the corporate business sector. See, e.g., the Fed's Financial Accounts of the United States at \url{https://www.federalreserve.gov/releases/z1/}.}
	
	In recent years, the macro-finance literature has further expanded on the macroeconomic role of housing, as documented in the surveys by \citet{DucaMuellbauerMurphy2021}, \citet{PiazzesiSchneider2016}, and \citet{DavisVanNieuwerburgh2015}, 
	highlighting the mutual interactions between house prices and debt.
	Given the key role of housing in collateralized lending, changes in house prices affect economic activity through their impact on credit to households and firms, the banking sector and the broader financial system. 
	They also affect output through direct wealth effects. These effects are typically strong, compared to those generated by other types of assets (e.g., stocks) due to the high marginal propensity to consume out of housing wealth, which is held to a large extent by indebted, low net-worth households (e.g., \citealt{MianSufi2015}).
	In good times, increases in house prices lead to relaxed credit constraints, more investment and consumption, more production and thus further upward pressures on prices. 
	Conversely, house price declines lead to tightened credit conditions, deleveraging, and downturns.
	Against this backdrop, a series of recent macro models with a housing sector (e.g., \citealt{FavilukisLudvigsonVanNieuwerburgh2017}; \citealt{JustinianoPrimiceriTambalotti2019}; \citealt{SommerSullivanVerbrugge2013}) explicitly investigate the relationship between aggregate economic activity and the price-rent ratio,	which is indicated by the asset pricing theory as a key valuation metric for housing, capturing the relationship between market values and fundamentals (e.g., \citealt{CampbellDavisGallinMartin2009}; \citealt{KishorMorley2015}).\footnote{Based on the standard present-value approach to asset pricing, house prices should equal discounted expected future rents (i.e., earnings on housing assets). Accordingly, the ratio should reflect expectations on housing returns (i.e., the discount factor, given by a risk-free interest rate plus a housing risk premium) and the growth rate of rents. 
		Like its stock-market counterpart, i.e., the price-earnings ratio, the price-rent ratio is commonly used to gauge whether assets are undervalued or overvalued, its long-term average serving as a benchmark to detect possible deviations of house prices from fundamental levels and to assess downside risks to house prices (e.g., \citealt{PhiliponnetTurrini2017}).
	}
	Overall, the literature suggests that, because of the role of housing in collateralized credit and the forward-looking nature of asset pricing, 
	the price-to-rent ratio combines information on current credit market conditions and expectations on future house prices (and thus the value of collateral). 
	In particular, a high price-rent ratio tends to be associated with lax credit constraints and expected house price depreciation.
	Such combination is likely to provide valuable predictive information on future business-cycle conditions, e.g., by signaling households' and firms' vulnerability to credit-tightening shocks and potential deleveraging. (Conversely, a low price-rent ratio may indicate that there is room for a future relaxation of credit constraints and an increase in housing wealth.)
	In an influential contribution, \citet{FavilukisLudvigsonVanNieuwerburgh2017} propose a general equilibrium model that helps rationalize such dual information provided by the price-rent ratio.
	During economic expansions, higher house prices (collateral values) allow households to borrow more easily, thus providing greater insurance against income risk. Accordingly, the housing risk premium decreases, pushing house prices further up. At the same time, higher housing demand prompts more residential investment, which lowers the expected growth rate of rents. As a result, a higher price-rent ratio can be only justified by lower expected housing returns (discount rates), in the form of future house price depreciation, rather than faster rental growth.\footnote{The model prediction that a high price-rent ratio forecasts lower house prices rather than high rental growth is consistent with the findings of a number of empirical papers, e.g., \citet{CampbellDavisGallinMartin2009} and \citet{KishorMorley2015}. Other theoretical models in which the price-rent ratio mostly reflects credit market conditions include \citet{JustinianoPrimiceriTambalotti2019}, \citet{GarrigaManuelliPeraltaAlva2019}, \citet{GreenwaldGuren2021}, and \citet{Chu2014}.  
	}
	The result that a high price-rent ratio tends to be associated with subsequent house price declines is also compatible with theoretical models of house price bubbles. \citet{AbrahamHendershott1996} propose a model in which extrapolative (i.e., backward-looking) expectations on house prices generate bubbles in good times. However, when prices become too high relative to fundamentals, expectations switch to negative, triggering a bust. \citet{DucaMuellbauerMurphy2011} find evidence of both time-varying credit constraints and extrapolative house price expectations as drivers of the U.S. price-to-rent ratio. 
	The predictive power of the ratio may also be related to its direct effect on home-ownership decisions (\citealt{SommerSullivan2018}): when the price-rent ratio increases, home-ownership becomes more expensive relative to renting. This may lead households to prefer renting over ownership, with negative effects on housing market activities and the business cycle in general.\footnote{See, e.g., the Fred Blog article ``Is the housing price-rent ratio a leading indicator?", available at \url{https://fredblog.stlouisfed.org/2018/09/is-the-housing-price-rent-ratio-a-leading-indicator/}. }
	
	Finally, \citet{CampbellShiller1998} suggest calculating ``cyclically-adjusted" valutation ratios, i.e., dividing asset prices by multi-year averages of earnings, as a way to smooth out noise in fundamentals and thus achieve more robust valuation. This motivates the use of the CAPR ratio instead of the simple price-rent ratio.

	\section{Conclusions}\label{sec:conclusions}
	
	In recent years, macroeconomic research has emphasized the role of financial conditions as key determinants of aggregate fluctuations. In particular, housing and debt cycles, building up in the background of business cycles, are now widely recognized to have profound and potentially disruptive effects on economic activity. 
	This paper provides new empirical results on the role of specific financial-cycle indicators for predicting U.S. GDP over medium-term horizons (1-5 years). 
	Based on a wide variety of methodologies applied to a high-dimensional dataset, we find that two ratios have a particularly strong relationship with economic activity over subsequent years, both combining information on financial conditions and economic fundamentals: the CAPR ratio, a robust valuation ratio for the housing market, and the NNBLI ratio, capturing the debt burden of noncorporate (small) firms.
	High (low) values of these ratios predict low (high) output growth over the medium term.
	Compared to composite measures of financial conditions, these indicators appear to offer more stable predictive information on GDP across different business-cycle phases and different time periods. 
	Also, their predictive ability appears to be consistent with macro-financial theories in which the interaction of housing market valuations and collateralized borrowing by firms and households represents a crucial transmission mechanism of economic fluctuations. 
	The results of the paper show that a careful selection of financial-cycle indicators provides substantial value added to multi-year forecasts of GDP. Small models that include the best indicators are able to outperform more sophisticated, high-dimensional models. 
	The CAPR and NNBLI ratios may thus be important tools for forecasting, economic analysis, and macro-prudential policy.

	\bibliography{references}

\begin{appendices}

\setcounter{table}{0}
\setcounter{figure}{0}
\renewcommand{\thetable}{A\arabic{table}}
\renewcommand{\thefigure}{A\arabic{figure}}

	\section*{Online Appendix. Tables and Figures}\label{sec:robustness}

	In this Appendix, we provide several tables and figures which contain results discussed in Section \ref{MAIN-sec:robustness} (``Extensions and robustness checks") of the paper.
	

	\vspace{10pt}
\begin{table}[H]
	\caption{High-dimensional models and forecast combinations: MSFEs for year-on-year GDP growth}
	\centering
	\scalebox{0.85}[0.85]{
		\begin{tabular}{c|ccc }
			\hline \hline 
			forecasting model/method	    & $h=4$      & $h=12$     & $h=20$     \\
			\hline
			
			LBVAR                                  & 0.82 & 0.88 & 0.81 \\
			LASSO VAR                              & 0.87 & 0.94 & 0.99 \\
			Factor model                           & 0.91 & 0.97 & 1.00 \\
			Forecast combination (equal   weights) & 0.93 & 1.00 & 1.00 \\
			Forecast combination (BMA   weights)   & 0.96 & 1.00 & 1.00 \\
			IMF                                    & 0.76 & 1.08 & 1.13\\
			\hline \hline 
		\end{tabular}
	}
	\subcaption*{\textit{Notes:} The table shows the mean squared forecast errors (MSFEs) for the $h$-quarter-ahead year-on-year GDP growth rate, relative to the benchmark AR, over the period 1990Q1-2017Q4. The forecasting models/methods considered are the same as in Table \ref{MAIN-tab:mse_hd_models} of the paper.}
	\label{tab:mse_hd_models_d4}
\end{table}


	\begin{table}[H]
		\centering
		\caption{Forecasts based on rolling-window estimates: MSFEs}%
		\scalebox{0.75}[0.75]{
			\begin{tabular}{l|ll|ll|ll}
				\hline \hline 
				& \multicolumn{2}{c|}{$h=4$}       & \multicolumn{2}{c|}{$h=12$} & \multicolumn{2}{c}{$h=20$}  \\
				\hline
				& \multicolumn{6}{c}{\textbf{ARDL}} \\
				\hline
				1  & \textbf{NNBTILQ027SBDIx} & 0.61 & \textbf{NNBTILQ027SBDIx} & 0.31 & pr              & 0.40 \\
				2  & cli             & 0.87 & IPCONGD         & 0.83 & \textbf{NNBTILQ027SBDIx} & 0.49 \\
				3  & AMDMUOx         & 0.96 & cli         	  & 0.86 & AMDMUOx         & 0.74 \\
				4  & TNWMVBSNNCBBDIx & 0.98 & TNWMVBSNNCBBDIx & 0.91 & CONSPIx         & 0.81 \\
				5  & TLBSNNCBBDIx    & 1.00 & DHLCRG3Q086SBEA & 0.93 & \textbf{capr}   & 0.82 \\
				6  & prfi_gdp    	 & 1.03 & IPNCONGD        & 0.96 & TNWMVBSNNCBBDIx & 0.97 \\
				7  & hpi             & 1.03 & CPIMEDSL        & 1.00 & cli             & 0.98 \\
				8  & PRFIx           & 1.08 & IPDCONGD        & 1.01 & CPIMEDSL 	   & 0.99 \\
				9  & MORTG10YRx      & 1.08 & EXUSUKx         & 1.02 & DHLCRG3Q086SBEA & 1.00 \\
				10 & USMINE          & 1.08 & DREQRG3Q086SBEA & 1.02 & PRFIx           & 1.02\\
				\hline
				& \multicolumn{6}{c}{\textbf{VAR}} \\
				\hline
				1  & \textbf{NNBTILQ027SBDIx}   & \textbf{0.72} & \textbf{NNBTILQ027SBDIx} & \textbf{0.40} & \textbf{NNBTILQ027SBDIx} & \textbf{0.72} \\
				2  & \textbf{capr}              & \textbf{0.96} & TNWMVBSNNCBBDIx & 0.80 & AMDMUOx         & 0.74 \\
				3  & cli               			& 0.97 & AMDMUOx         & 0.84 & \textbf{capr}            & \textbf{0.95} \\
				4  & CPIAUCSL              		& 1.01 & cli             & 0.84 & cli             & 0.95 \\
				5  & AMDMUOx            		& 1.06 & \textbf{capr}   & \textbf{1.02} & TNWMVBSNNCBBDIx & 1.00 \\
				6  & PRFIx          			& 1.09 & USMINE          & 1.10 & BUSLOANSx       & 1.10 \\
				7  & TNWMVBSNNCBBDIx            & 1.10 & CPIAUCSL        & 1.12 & USTPU           & 1.18 \\
				8  & WPU0561           			& 1.10 & B021RE1Q156NBEA & 1.18 & HOUST5F         & 1.21 \\
				9  & B021RE1Q156NBEA  			& 1.12 & PRFIx           & 1.20 & IPMAT           & 1.25 \\
				10 & hpi           				& 1.12 & HOUST5F         & 1.23 & S\&P: PE ratio  & 1.25 \\
				\hline \hline 
			\end{tabular}
		}
		\subcaption*{\textit{Notes}: MSFEs for the $h$-quarter-ahead (cumulative) GDP growth rate, relative to the benchmark AR, calculated over the period 1990Q1-2017Q4. All models are estimated on rolling windows of size 60 quarters. See Table \ref{MAIN-tab:redux} of the paper and \citet{McCrackenNgFREDQD} for a description of the variables.}
		\label{tab:mse_ardl_var_rolling} 
	\end{table}


	\begin{table}[H]
		\centering
		\caption{$R^2$ of ARDL models estimated on sub-samples}%
		\scalebox{0.7}[0.7]{
			\begin{tabular}{l|ll|ll|ll}
				\hline \hline 
				& \multicolumn{2}{c|}{$h=4$}       & \multicolumn{2}{c|}{$h=12$} & \multicolumn{2}{c}{$h=20$}  \\
				\hline
				& \multicolumn{6}{c}{\textbf{(a) sample: excluding 2007Q3-2009Q2 (GFC)}} \\
				\hline
				1  & cli             & 0.36 & \textbf{capr}            & 0.61 & \textbf{capr}            & 0.68 \\
				2  & DOTSRG3Q086SBEA & 0.34 & pr              & 0.56 & pr              & 0.65 \\
				3  & fedfunds        & 0.34 & \textbf{NNBTILQ027SBDIx} & 0.55 & \textbf{NNBTILQ027SBDIx} & 0.59 \\
				4  & CP3M            & 0.33 & AMDMUOx         & 0.52 & LIABPIx         & 0.58 \\
				5  & IPCONGD         & 0.33 & cli             & 0.51 & mortg_inc       & 0.58 \\
				6  & CP3M            & 0.31 & mortg_inc       & 0.42 & ISRATIOx        & 0.52 \\
				7  & TB3MS           & 0.31 & ISRATIOx        & 0.41 & cred_gdp        & 0.52 \\
				8  & MORTGAGE30US    & 0.31 & LIABPIx         & 0.41 & cli             & 0.50 \\
				9  & \textbf{capr}            & 0.31 & cred_gdp        & 0.39 & NWPIx           & 0.47 \\
				10 & TB6MS           & 0.30 & AAA             & 0.38 & AMDMUOx         & 0.47 \\
				\hline
				& \multicolumn{6}{c}{\textbf{(b) sample: up to 2007Q2}} \\
				\hline
				1  & fedfunds        & 0.38 & UEMPMEAN        & 0.59 & UEMPMEAN      & 0.78 \\
				2  & IMPGSC1         & 0.38 & hpi             & 0.57 & pr            & 0.68 \\
				3  & MORTGAGE30US    & 0.37 & prfi_gdp        & 0.57 & UNRATELTx     & 0.63 \\
				4  & CP3M            & 0.36 & CPIAUCSL        & 0.56 & \textbf{capr}          & 0.59 \\
				5  & TB3MS           & 0.36 & PCECTPI         & 0.55 & hpi           & 0.58 \\
				6  & DOTSRG3Q086SBEA & 0.36 & UNRATELTx       & 0.54 & USGOVT        & 0.57 \\
				7  & prfi_gdp        & 0.35 & CUSR0000SA0L5   & 0.54 & CONSPIx       & 0.57 \\
				8  & BAA             & 0.35 & IMPGSC1         & 0.54 & mortg         & 0.57 \\
				9  & TB6M3Mx         & 0.35 & \textbf{capr}            & 0.53 & AWHNONAG      & 0.56 \\
				10 & TB6M3Mx         & 0.35 & DHLCRG3Q086SBEA & 0.53 & CES9093000001 & 0.53\\
				\hline
				& \multicolumn{6}{c}{\textbf{(c) sample: ``Great Moderation", 1983Q1-2007Q2}} \\
				\hline				
				1  & TLBSNNCBBDIx    & 0.53 & \textbf{NNBTILQ027SBDIx} & 0.62 & UEMPMEAN      & 0.81 \\
				2  & \textbf{NNBTILQ027SBDIx} & 0.53 & \textbf{capr}            & 0.61 & \textbf{capr}          & 0.74 \\
				3  & CUSR0000SAS     & 0.52 & DHLCRG3Q086SBEA & 0.59 & UNRATELTx     & 0.73 \\
				4  & USGOVT          & 0.46 & OUTMS           & 0.59 & OUTMS         & 0.69 \\
				5  & cred_gdp        & 0.46 & UEMPMEAN        & 0.57 & REALLNx       & 0.69 \\
				6  & LNS14000012     & 0.46 & GDPCTPI         & 0.57 & USGOVT        & 0.69 \\
				7  & CPILFESL        & 0.46 & PCEPILFE        & 0.55 & pr            & 0.67 \\
				8  & DHUTRG3Q086SBEA & 0.44 & USGOVT          & 0.54 & CES9093000001 & 0.64 \\
				9  & UNRATELTx       & 0.43 & IMPGSC1         & 0.53 & mortg         & 0.64 \\
				10 & USFIRE          & 0.43 & IPDBS           & 0.52 & IMPGSC1       & 0.62\\
				\hline \hline 
			\end{tabular}
		}
		\subcaption*{\textit{Notes:} For each $h$, the dependent variable is GDP growth over $h$ quarters. Please refer to Section \ref{MAIN-subsec:subsamples} of the paper for details.}
		\label{tab:r2_subsamples}
	\end{table}

	\begin{table}[H]
		\centering
		\caption{ARDL forecasts: MSFEs calculated on sub-periods}%
		\scalebox{0.75}[0.75]{
			\begin{tabular}{l|ll|ll|ll}
				\hline \hline 
				& \multicolumn{2}{c|}{$h=4$}       & \multicolumn{2}{c|}{$h=12$} & \multicolumn{2}{c}{$h=20$} \\
				\hline
				& \multicolumn{6}{c}{\textbf{pre-Crisis period: 1990Q1-2007Q2}} \\
				\hline
				1  & \textbf{NNBTILQ027SBDIx} & 0.61 & \textbf{NNBTILQ027SBDIx} & 0.46 & \textbf{NNBTILQ027SBDIx} & 0.46 \\
				2  & FPIx            & 0.83 & LNS14000012     & 0.77 & \textbf{capr}            & 0.57 \\
				3  & S\&P div. yield   & 0.89 & IMPGSC1         & 0.81 & VXOCLSX         & 0.60 \\
				4  & CPILFESL        & 0.90 & UNRATELTx       & 0.82 & LNS14000012     & 0.62 \\
				5  & CPIMEDSL        & 0.90 & PERMITNE        & 0.83 & UNRATELTx       & 0.65 \\
				6  & CUSR0000SAS     & 0.90 & HOUSTNE         & 0.87 & hpi             & 0.70 \\
				7  & OUTNFB          & 0.90 & IPCONGD         & 0.89 & HWIURATIOx      & 0.78 \\
				8  & HNOREMQ027Sx    & 0.90 & TLBSNNCBBDIx    & 0.90 & UEMP27OV        & 0.79 \\
				9  & LNS14000012     & 0.90 & \textbf{capr}            & 0.91 & AWHNONAG        & 0.80 \\
				10 & PERMITS         & 0.91 & DREQRG3Q086SBEA & 0.91 & UNRATE          & 0.80\\
				\hline 
				& \multicolumn{6}{c}{\textbf{Crisis period: 2007Q3-2009Q2}} \\
				\hline
				1  & \textbf{NNBTILQ027SBDIx} & 0.28 & USSTHPI         & 0.14 & \textbf{capr}         & 0.20 \\
				2  & \textbf{capr}            & 0.29 & prfi_gdp        & 0.18 & hpi          & 0.33 \\
				3  & PRFIx           & 0.38 & hpi             & 0.20 & USSTHPI      & 0.34 \\
				4  & cli             & 0.42 & HNOREMQ027Sx    & 0.32 & HNOREMQ027Sx & 0.40 \\
				5  & hpi             & 0.43 & SPCS10RSA       & 0.32 & TNWBSNNBx    & 0.42 \\
				6  & NWPIx           & 0.44 & pr              & 0.32 & prfi_gdp     & 0.46 \\
				7  & prfi_gdp        & 0.47 & \textbf{capr}   & 0.33 & OUTMS        & 0.48 \\
				8  & NNBTASQ027Sx    & 0.48 & TNWBSNNBx       & 0.39 & pr           & 0.49 \\
				9  & mortg           & 0.52 & \textbf{NNBTILQ027SBDIx} & 0.40 & TWEXMMTH     & 0.53 \\
				10 & B020RE1Q156NBEA & 0.53 & NNBTASQ027Sx    & 0.41 & mortg        & 0.56\\
				\hline 
				& \multicolumn{6}{c}{\textbf{post-Crisis period: 2009Q3-2017Q4}} \\
				\hline
				1  & IPNCONGD        & 0.67 & \textbf{NNBTILQ027SBDIx} & 0.35 & \textbf{capr}            & 0.31 \\
				2  & pr              & 0.68 & \textbf{capr}            & 0.59 & pr              & 0.48 \\
				3  & PCECC96         & 0.71 & cli             & 0.63 & \textbf{NNBTILQ027SBDIx} & 0.49 \\
				4  & cred_gdp        & 0.73 & ISRATIOx        & 0.79 & cred_gdp        & 0.70 \\
				5  & LIABPIx         & 0.76 & TWEXMMTH        & 0.85 & LIABPIx         & 0.82 \\
				6  & bci             & 0.76 & LIABPIx         & 0.89 & ISRATIOx        & 0.85 \\
				7  & B021RE1Q156NBEA & 0.76 & cred_gdp        & 0.91 & B021RE1Q156NBEA & 0.86 \\
				8  & IPCONGD         & 0.77 & mortg_inc       & 0.91 & cli             & 0.89 \\
				9  & TWEXMMTH        & 0.80 & NWPIx           & 0.98 & MORTGAGE30US    & 0.99 \\
				10 & ISRATIOx        & 0.80 & AMDMUOx         & 1.00 & TWEXMMTH        & 1.02\\
				\hline \hline 
			\end{tabular}
		}
		\subcaption*{\textit{Notes}: MSFEs for the $h$-quarter (cumulative) GDP growth rate, relative to the benchmark AR model. Please refer to Section \ref{MAIN-subsec:subsamples} of the paper for details.}
		\label{tab:mse_ardl_subsamples} 
	\end{table}

	\begin{table}[H]
		\centering
		\caption{VAR forecasts: MSFEs calculated on sub-periods}%
		\scalebox{0.75}[0.75]{
			\begin{tabular}{l|ll|ll|ll}
				\hline \hline 
				& \multicolumn{2}{c|}{$h=4$}       & \multicolumn{2}{c|}{$h=12$} & \multicolumn{2}{c}{$h=20$} \\
				\hline
				& \multicolumn{6}{c}{\textbf{pre-Crisis period: 1990Q1-2007Q2}} \\
				\hline
				1  & \textbf{NNBTILQ027SBDIx} 	& 0.59 & \textbf{NNBTILQ027SBDIx} 	& 0.63 & UNRATELTx       			& 0.66 \\
				2  & prfi_gdp        			& 0.81 & IMPGSC1         			& 0.72 & LNS14000012     			& 0.73 \\
				3  & OUTNFB          			& 0.84 & LNS14000012     			& 0.75 & REALLNx         			& 0.73 \\
				4  & S\&P: div. yield			& 0.84 & UNRATELTx       			& 0.80 & TB3SMFFM        			& 0.74 \\
				5  & LNS14000012      			& 0.85 & HOUSTNE         			& 0.81 & IMPGSC1         			& 0.75 \\
				6  & PERMITS	    			& 0.88 & CPIMEDSL        			& 0.83 & HNOREMQ027Sx    			& 0.78 \\
				7  & TLBSNNCBBDIx      			& 0.88 & IPNCONGD        			& 0.87 & \textbf{NNBTILQ027SBDIx} 	& 0.79 \\
				8  & CPIMEDSL          			& 0.90 & PERMITNE        			& 0.88 & VXOCLSX         			& 0.79 \\
				9  & BAA            			& 0.90 & OUTNFB          			& 0.89 & prfi_gdp        			& 0.80 \\
				10 & OUTBS          			& 0.90 & GPDIC1          			& 0.90 & GPDIC1          			& 0.82\\
				\hline 
				& \multicolumn{6}{c}{\textbf{Crisis period: 2007Q3-2009Q2}} \\
				\hline
				1  & PRFIx           			& 0.20 & pr              			& 0.15 & HNOREMQ027Sx & 0.48 \\
				2  & \textbf{capr}            	& 0.23 & \textbf{capr}            	& 0.18 & ISRATIOx     & 0.55 \\
				3  & \textbf{NNBTILQ027SBDIx} 	& 0.31 & \textbf{NNBTILQ027SBDIx} 	& 0.32 & prfi_gdp     & 0.57 \\
				4  & cli             			& 0.35 & prfi_gdp        			& 0.35 & TNWBSNNBx    & 0.59 \\
				5  & hpi             			& 0.36 & ISRATIOx        			& 0.36 & REVOLSLx     & 0.63 \\
				6  & pr              			& 0.40 & HNOREMQ027Sx    			& 0.36 & USSTHPI      & 0.63 \\
				7  & PERMITW         			& 0.44 & USSTHPI         			& 0.40 & TWEXMMTH     & 0.66 \\
				8  & HOUST           			& 0.45 & AMDMUOx         			& 0.45 & TLBSHNOx     & 0.69 \\
				9  & HOUSTS          			& 0.46 & TB3MS           			& 0.53 & UNRATESTx    & 0.70 \\
				10 & PERMITMW        			& 0.46 & NNBTASQ027Sx    			& 0.53 & PCESVx       & 0.70\\
				\hline 
				& \multicolumn{6}{c}{\textbf{post-Crisis period: 2009Q3-2017Q4}} \\
				\hline
				1  & A014RE1Q156NBEA & 0.70 & pr              			& 0.15 & pr              & 0.07 \\
				2  & PCECC96         & 0.74 & \textbf{capr}            	& 0.15 & \textbf{capr}            & 0.09 \\
				3  & S\&P: div. yield& 0.78 & \textbf{NNBTILQ027SBDIx} 	& 0.18 & \textbf{NNBTILQ027SBDIx} & 0.35 \\
				4  & pr              & 0.80 & cli             			& 0.51 & ISRATIOx        & 0.37 \\
				5  & ISRATIOx        & 0.80 & ISRATIOx        			& 0.53 & TWEXMMTH        & 0.60 \\
				6  & PERMITW         & 0.81 & TWEXMMTH        			& 0.62 & B021RE1Q156NBEA & 0.60 \\
				7  & cape            & 0.81 & B021RE1Q156NBEA 			& 0.65 & NWPIx           & 0.62 \\
				8  & IPCONGD         & 0.81 & UNRATESTx       			& 0.66 & USMINE          & 0.66 \\
				9  & UNRATESTx       & 0.81 & NWPIx           			& 0.66 & cli             & 0.70 \\
				10 & EXCAUSx         & 0.84 & AMDMUOx         			& 0.69 & UNRATESTx       & 0.71 \\
				\hline \hline
			\end{tabular}
		}
		\subcaption*{\textit{Notes}: MSFEs for the $h$-quarter (cumulative) GDP growth rate, relative to the benchmark AR model. Please refer to Section \ref{MAIN-subsec:subsamples} of the paper for details.}
		\label{tab:mse_var_subsamples} 
	\end{table}

	\begin{figure}[H]
		\caption{Forecasting other recessions: CAPR and NNBLI}
		\centering
		\hspace{10pt}
		\begin{subfigure}[b]{0.35\textwidth}
			\caption{1990-1991 recession}
			\centering
			\includegraphics[width=\textwidth]{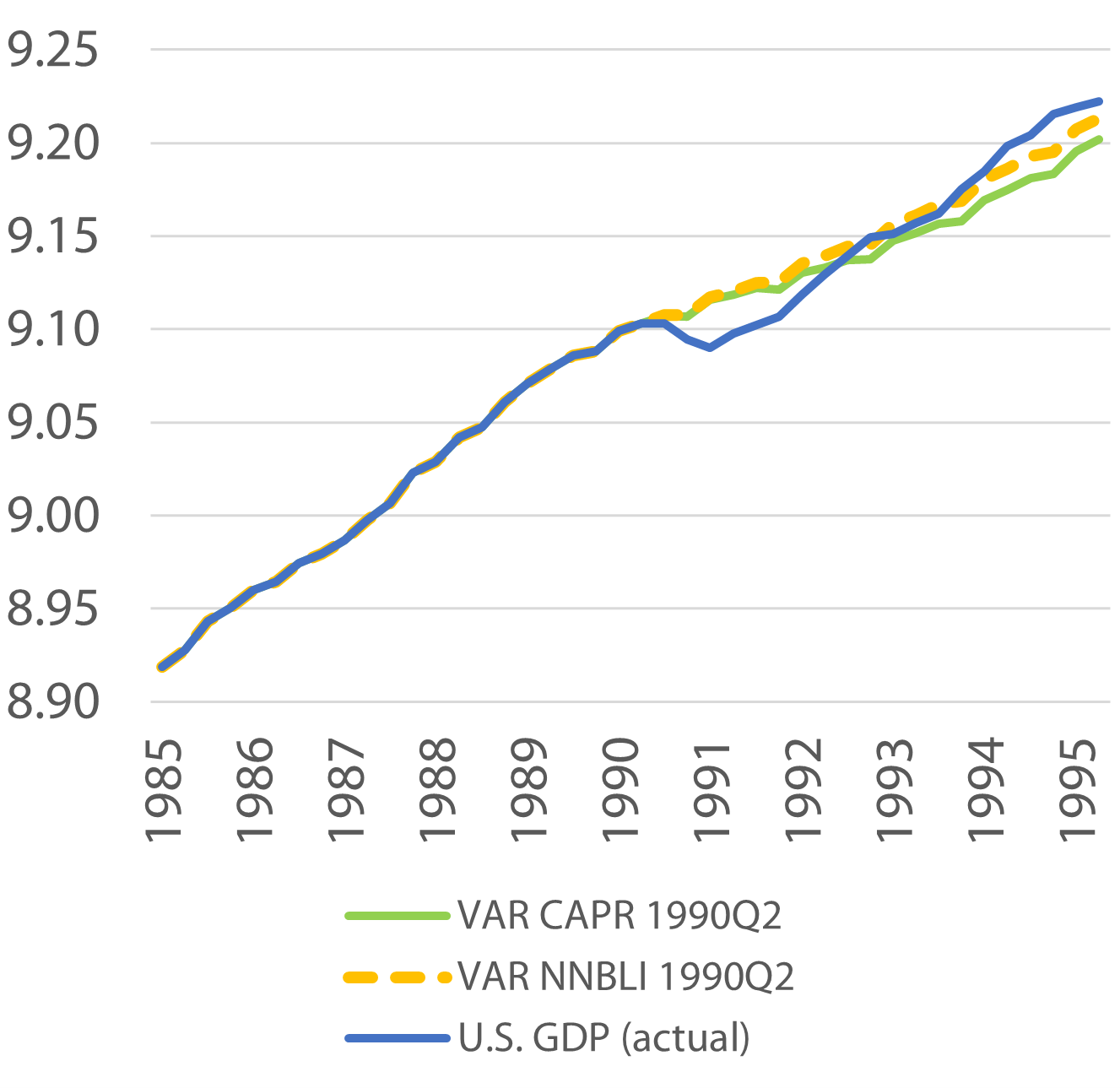}
		\end{subfigure}
		\hspace{10pt}
		\begin{subfigure}[b]{0.35\textwidth}
			\caption{2001 recession}
			\centering
			\includegraphics[width=\textwidth]{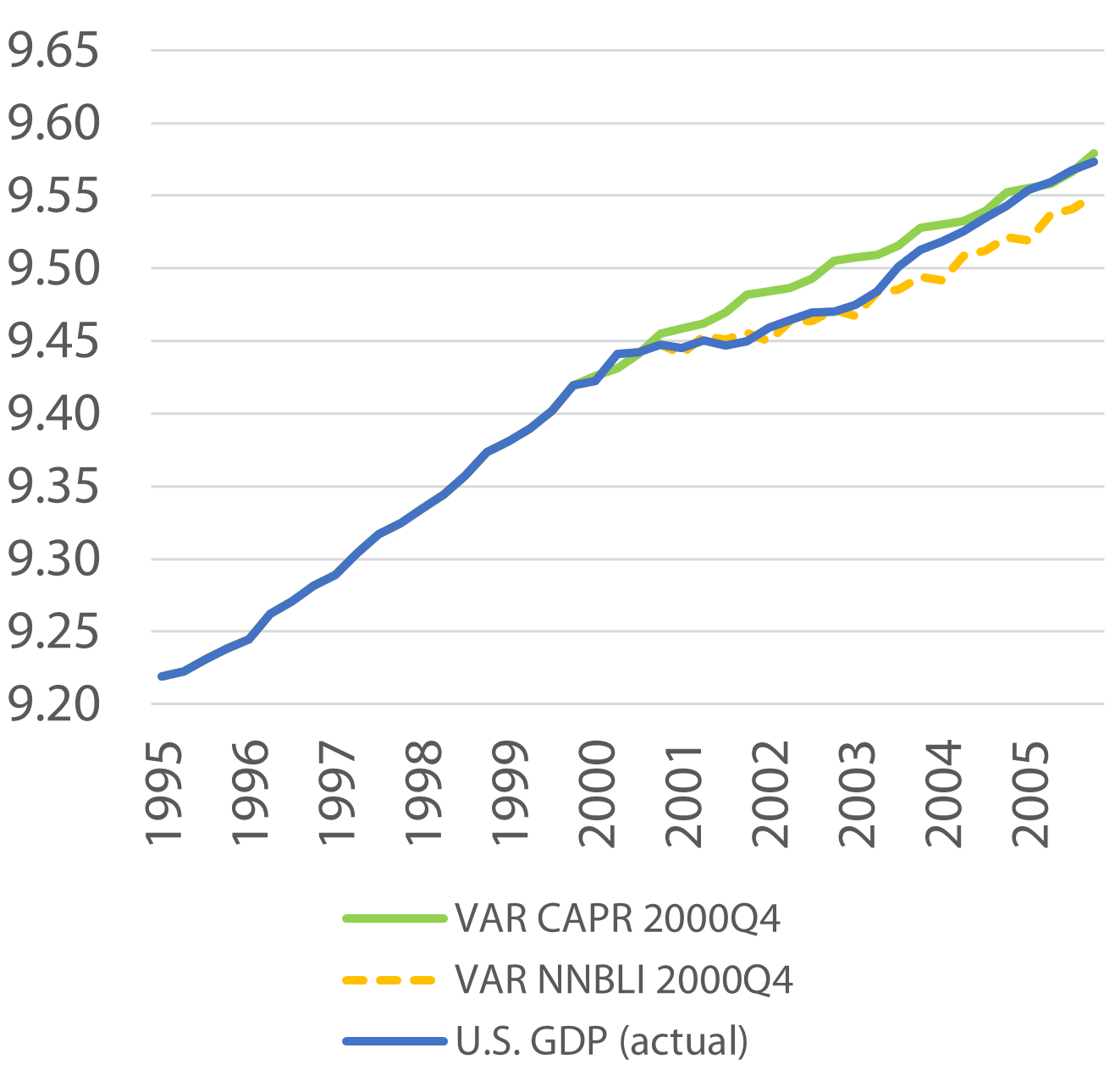}
		\end{subfigure}
		\subcaption*{\textit{Notes:} This figure shows forecasts of U.S. GDP produced by bivariate VAR models using either CAPR or NNBLI. In panel (a), forecasts are generated in 1990Q2 using models estimated on the sample 1967Q1-1990Q2.  In panel (b), they are generated in 2000Q4 using models estimated on the sample 1967Q1-2000Q4.}
		\label{fig:forecast_recessions}
	\end{figure}


	\begin{landscape}
		
		\begin{figure}[H]
			\caption{Comparing forecast accuracy under instability: the Giacomini-Rossi (2010) test}
			\centering
			\begin{subfigure}[b]{0.27\textwidth}
				\caption*{\tiny ($h=4$) NNBLI vs. LBVAR}
				\centering
				\includegraphics[width=\textwidth]{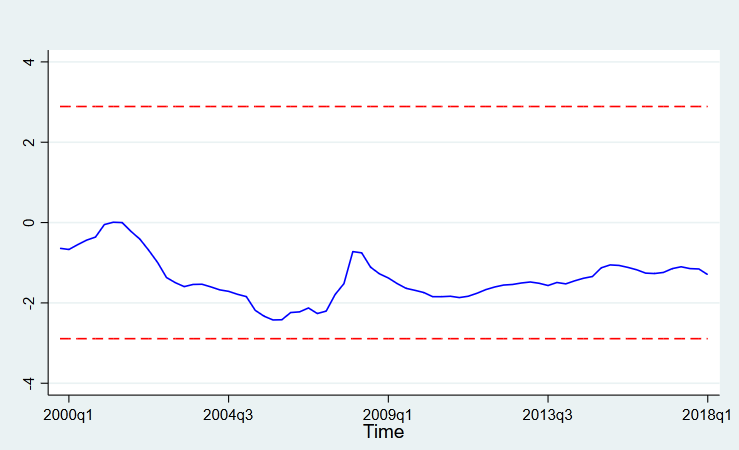}
			\end{subfigure}
			\begin{subfigure}[b]{0.27\textwidth}
				\caption*{\tiny ($h=4$) LASSO VAR}
				\centering
				\includegraphics[width=\textwidth]{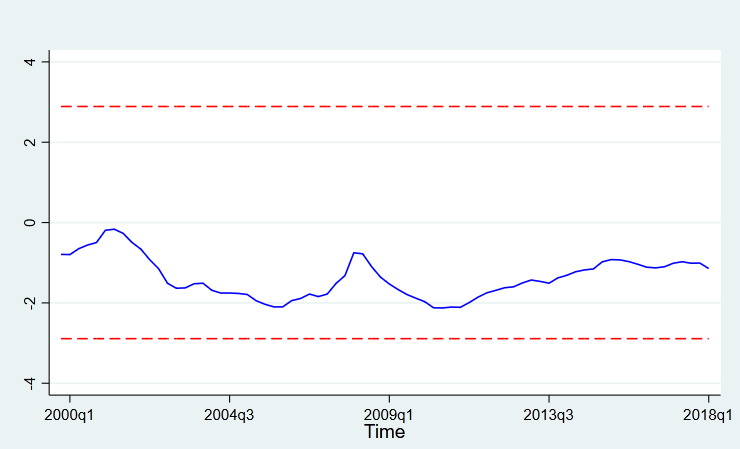}
			\end{subfigure}
			\begin{subfigure}[b]{0.27\textwidth}
				\caption*{\tiny ($h=4$) NNBLI vs. factor model}
				\centering
				\includegraphics[width=\textwidth]{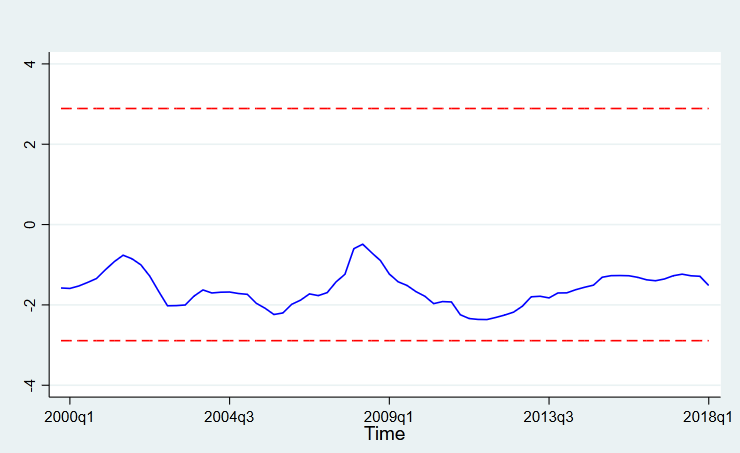}
			\end{subfigure}
			\begin{subfigure}[b]{0.27\textwidth}
				\caption*{\tiny ($h=4$) NNBLI vs. for. comb.}
				\centering
				\includegraphics[width=\textwidth]{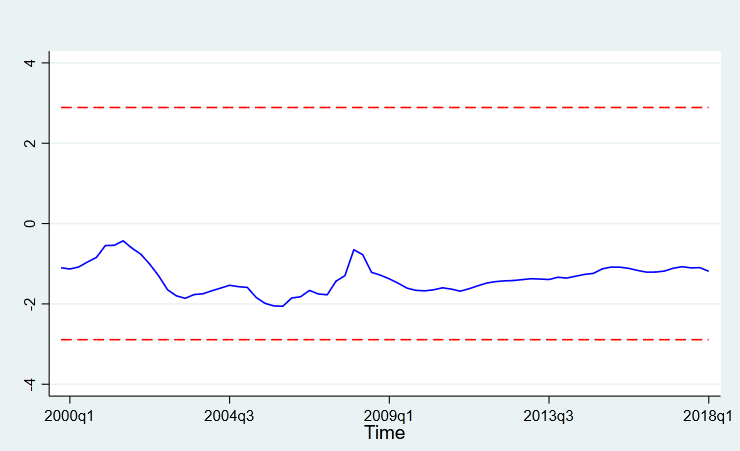}
			\end{subfigure}
			\begin{subfigure}[b]{0.27\textwidth}
				\caption*{\tiny ($h=4$) NNBLI vs. IMF}
				\centering
				\includegraphics[width=\textwidth]{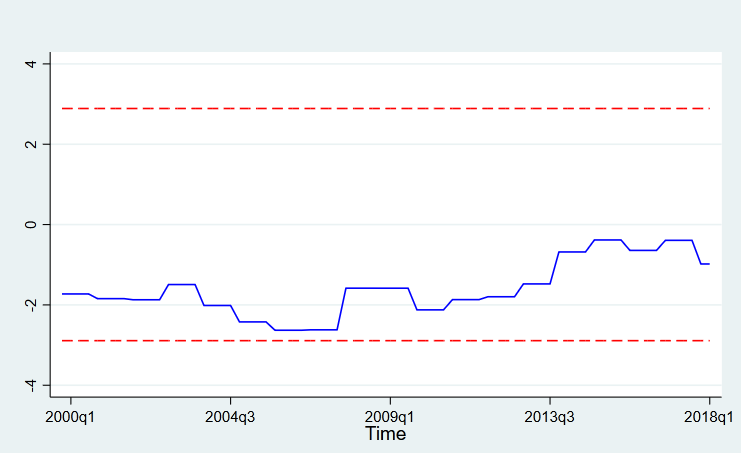}
			\end{subfigure}
			\begin{subfigure}[b]{0.27\textwidth}
				\caption*{\tiny ($h=12$) NNBLI vs. LBVAR}
				\centering
				\includegraphics[width=\textwidth]{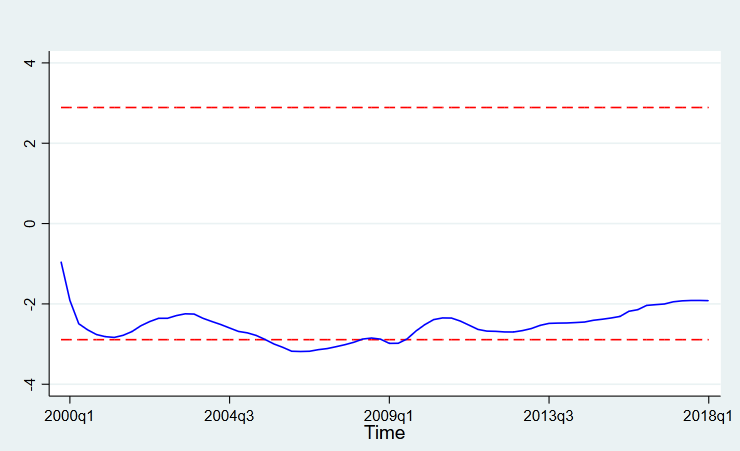}
			\end{subfigure}
			\begin{subfigure}[b]{0.27\textwidth}
				\centering
				\caption*{\tiny ($h=12$) NNBLI vs. LASSO VAR}
				\includegraphics[width=\textwidth]{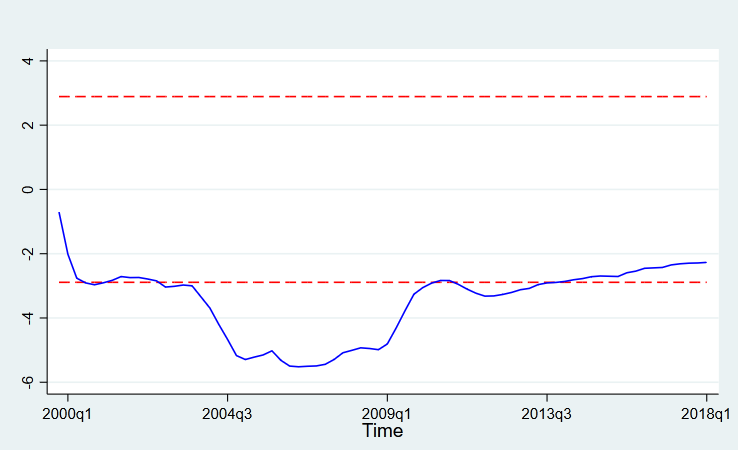}
			\end{subfigure}
			\begin{subfigure}[b]{0.27\textwidth}
				\centering
				\caption*{\tiny ($h=12$) NNBLI vs. factor model}
				\includegraphics[width=\textwidth]{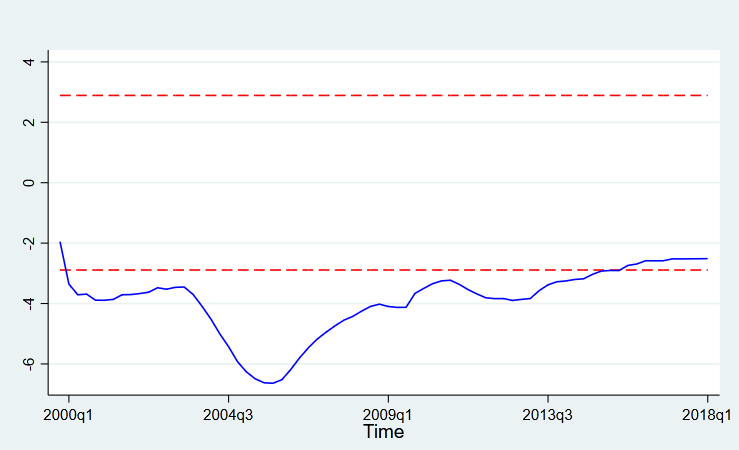}
			\end{subfigure}
			\begin{subfigure}[b]{0.27\textwidth}
				\centering
				\caption*{\tiny ($h=12$) NNBLI vs. for. comb.}
				\includegraphics[width=\textwidth]{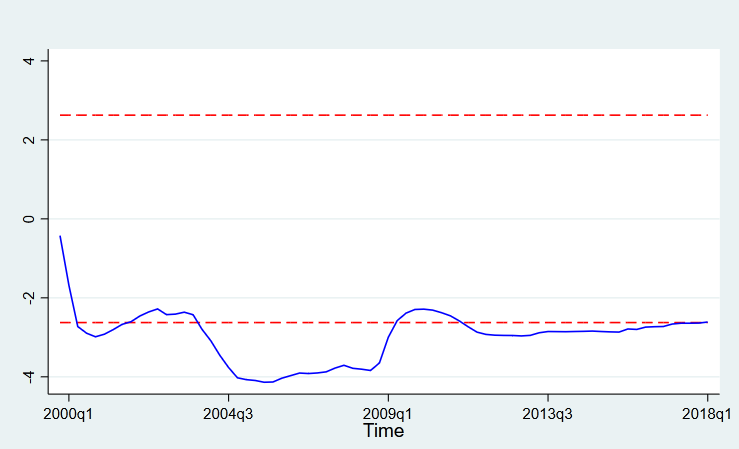}
			\end{subfigure}
			\begin{subfigure}[b]{0.27\textwidth}
				\centering
				\caption*{\tiny ($h=12$) NNBLI vs. IMF}
				\includegraphics[width=\textwidth]{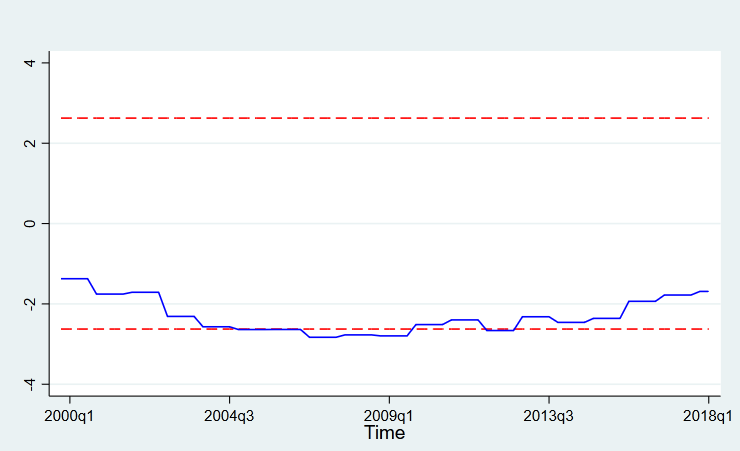}
			\end{subfigure}
			\begin{subfigure}[b]{0.27\textwidth}
				\centering
				\caption*{\tiny ($h=20$) CAPR vs. LBVAR}
				\includegraphics[width=\textwidth]{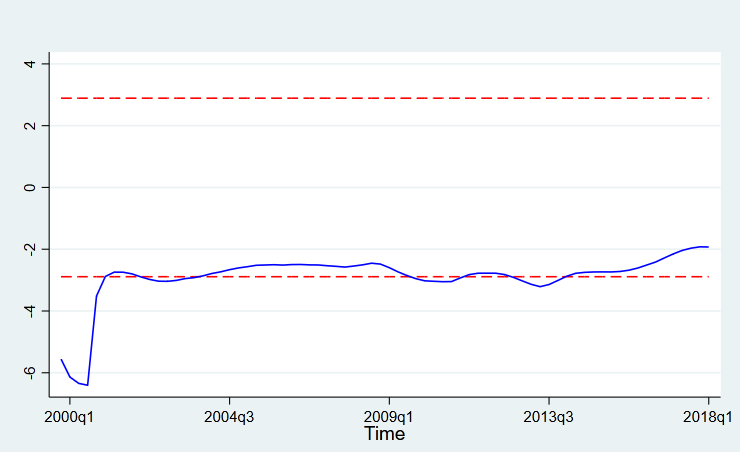}
			\end{subfigure}
			\begin{subfigure}[b]{0.27\textwidth}
				\centering
				\caption*{\tiny ($h=20$) CAPR vs. LASSO VAR}
				\includegraphics[width=\textwidth]{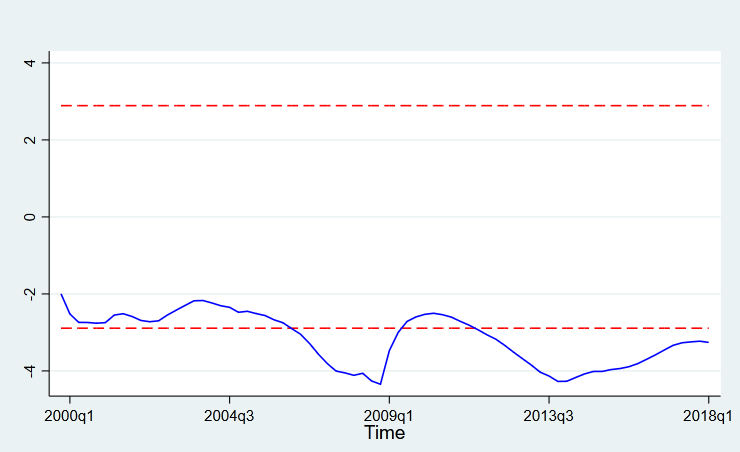}
			\end{subfigure}
			\begin{subfigure}[b]{0.27\textwidth}
				\centering
				\caption*{\tiny ($h=20$) CAPR vs. factor model}
				\includegraphics[width=\textwidth]{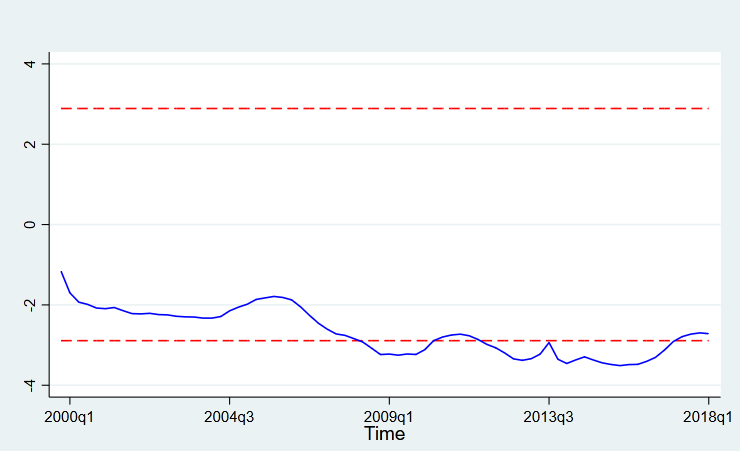}
			\end{subfigure}
			\begin{subfigure}[b]{0.27\textwidth}
				\centering
				\caption*{\tiny ($h=20$) CAPR vs. for. comb.}
				\includegraphics[width=\textwidth]{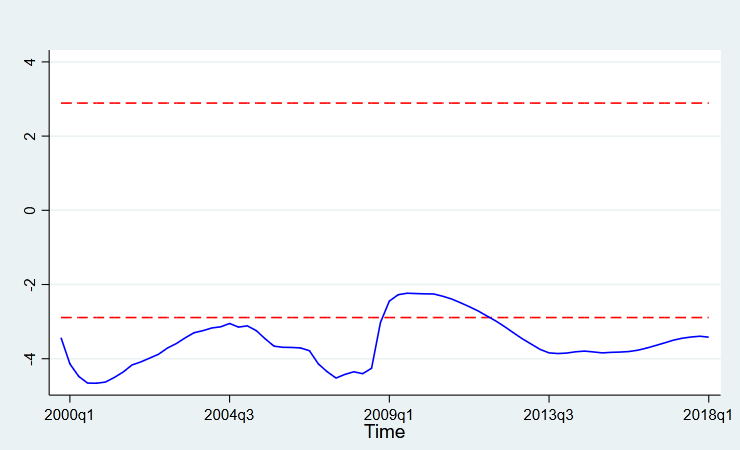}
			\end{subfigure}
			\begin{subfigure}[b]{0.27\textwidth}
				\centering
				\caption*{\tiny ($h=20$) CAPR vs. IMF}
				\includegraphics[width=\textwidth]{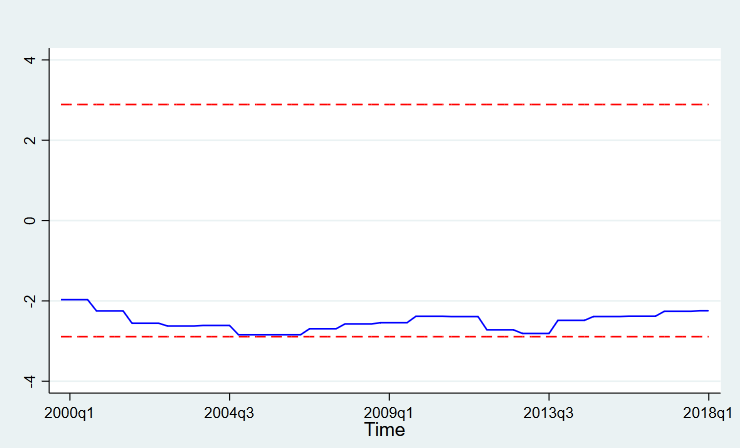}
			\end{subfigure}
			\subcaption*{\textit{Notes:} The figure shows the sequence of loss differences over time (continuous blue line) between forecasts produced by one-predictor models using NNBLI or CAPR and forecasts by competing high-dimensional models, along with the 5\% critical values of the \citet{GiacominiRossi2010} test (dashed red lines). A loss difference crossing the lower dashed line indicates that forecasts based on NNBLI or CAPR significantly outperform the competing forecasts.}
			\label{fig:gr2010test}
		\end{figure}
		\vfill
	\end{landscape}


	\begin{landscape}
		\begin{table}[h]
			\centering
			\caption{Comparing variable selection methods}%
			\scalebox{0.53}[0.6]{
				\begin{tabular}{|Hlll|lll|lll|l|}
					\hline \hline 
					&  \multicolumn{3}{c|}{\textbf{LASSO} ($\lambda=0.005$)}              &  \multicolumn{3}{c|}{\textbf{Shrinkage (``horseshoe") prior}}  &  \multicolumn{3}{c|}{\textbf{Random Forest}}     & \textbf{LASSO VAR} (GDP equation)       \\
					\hline
					& $h=4  $               &$ h=12$                 & $h=20   $              & $h=4  $           & $h=12    $        & $h=20 $    & $h=4  $           & $h=12    $        & $h=20 $        & \\
					\hline
					1  & DOTSRG3Q086SBEA 				& \textbf{capr}    & AHETPIx         & LNS14000012     			& AWHMAN          			& BUSLOANSx      				 & pr     		 & \textbf{capr}            & \textbf{capr}         & bci             \\
					2  & MORTGAGE30US   			 	& AHETPIx         & AMDMUOx         & \textbf{NNBTILQ027SBDIx} & CLI            			 & EXJPUSx        		 		& AAAFFM         		 & pr       & pr    & CLAIMSx         \\
					3  & \textbf{NNBTILQ027SBDIx} 		& AMDMUOx         & \textbf{capr}  & TNWMVBSNNCBBDIx 			& HWIx            				& HWIx           		 	& \textbf{NNBTILQ027SBDIX} & AMDMUOX         & cred\_gdp      & cli             \\
					4  & PCESVx          				& CES9091000001   & CES3000000008x  & UMCSENTx       			 & \textbf{NNBTILQ027SBDIx} 	& \textbf{NNBTILQ027SBDIx}	 & cred\_gdp        			& \textbf{NNBTILQ027SBDIx} & mortg\_inc     & DODGRG3Q086SBEA \\
					5  & PERMITMW       				 & DOTSRG3Q086SBEA & cli             & VXOCLSx       			  & TNWBSNNBBDIx    			& S\&P: div. yield   				& AMBSLREALX      		& cred\_gdp        & hpi           & DONGRG3Q086SBEA \\
					6  & T5YFFM          				& DSERRG3Q086SBEA & cred\_gdp        &                			 & TNWMVBSNNCBBDIx 				& TLBSNNCBBDIx    				& T5YFFM          		& mortg\_inc       & prfi\_gdp     & GDPC1           \\
					7  & USMINE         				 & HWIX            & LNS13023705     &               			  & UMCSENTx       			 & TNWMVBSNNCBBDIx					 & mortg\_inc      		 & USMINE          & NWPIx        & mortg           \\
					8  &                				 & IPNCONGD        & pip_inc         &                		&                 			&                 					& \textbf{capr}            & GPDICTPI 	  	 & HNOREMQ027Sx & \textbf{NNBTILQ027SBDIx} \\
					9  &                			 	& USMINE          & USMINE          &                 		&                			 &                							 &                 & TCU 			   &  TNWBSNNBx   & OPHNFB          \\
					10 &                			 	& B020RE1Q156NBEA & COMPRNFB        &                 		&                			 &                							 &                 & prfi\_gdp        &              & PCECC96         \\
					11 &                				 & FEDFUNDS        & EXJPUSx         &                 		&                 			&                							 &                 & TB3MS           &              & PCESVx          \\
					12 &               			& \textbf{NNBTILQ027SBDIx} & GFDEBTNx 		&                 		&                			 &                							 &                 & HWIx            &              & PERMIT          \\
					13 &                				 & UMCSENTx        & M2REALx         &                 		&                 			&                							 &                 &                 &             			 & PERMITMW        \\
					14 &                 					&                 & \textbf{NNBTILQ027SBDIx} &         &               				  &               							  &                 &                 &              			& PRFIx           \\
					15 &                					 &                 & REVOLSLx        &                 &                			 &                							 &                 &                 &              			& S\&P: indust.     \\
					16 &                					 &                 & TWEXMMTH        &                 &                				 &                						 &                 &                 &              			& S\&P: div. yield   \\
					17 &                					 &                 &                 &                 &               					  &                						 &                 &                 &              			& UEMPLT5         \\
					18 &                					 &                 &                 &                 &                					 &                 						&                 &                 &              			& UMCSENTx       \\
					\hline \hline 
				\end{tabular}
			\hfill*
			}
			\subcaption*{\textit{Notes:} The table shows the list of predictors of GDP selected by different variables-selection methods (see Section \ref{MAIN-subsec:variable_selection} of the paper for details). For each $h$, the dependent variable is GDP growth over $h$ quarters.}
			\label{tab:variable_selection}
		\end{table}
	\end{landscape}


	\begin{table}[H]
		\centering
		\caption{Forecasts using real-time data: MSFEs}%
		\scalebox{0.75}[0.75]{
			\begin{tabular}{l|ll|ll|ll}
				\hline \hline 
				& \multicolumn{2}{c|}{$h=4$}       & \multicolumn{2}{c|}{$h=12$} & \multicolumn{2}{c}{$h=20$}  \\
				\hline 
				& \multicolumn{6}{c}{\textbf{ARDL}} \\
				\hline
1  & \textbf{capr}            & 0.69 & pr              & 0.54 & pr              & 0.23 \\
2  & \textbf{NNBTILQ027SBDIx} & 0.70 & \textbf{capr}            & 0.59 & \textbf{capr}            & 0.26 \\
3  & pr              & 0.72 & \textbf{NNBTILQ027SBDIx} & 0.62 & \textbf{NNBTILQ027SBDIx} & 0.76 \\
4  & HOUSTW          & 0.83 & AMDMUOx         & 0.96 & ISRATIOx        & 0.81 \\
5  & PERMITW         & 0.84 & AAA             & 0.99 & cli             & 0.96 \\
6  & PERMIT          & 0.86 & gs10            & 0.99 & IPCONGD         & 1.09 \\
7  & DPCERA3M086SBEA & 0.88 & USTRADE         & 1.00 & EXUSUKx         & 1.11 \\
8  & PERMITMW        & 0.88 & GS5             & 1.00 & UMCSENTx        & 1.12 \\
9  & prfi\_gdp       & 0.91 & BAA             & 1.01 & IPDCONGD        & 1.12 \\
10 & PERMITS         & 0.93 & DPCERA3M086SBEA & 1.06 & IPNMAT          & 1.13\\
				\hline 
				& \multicolumn{6}{c}{\textbf{VAR}} \\
				\hline
1  & pr               & 0.66 & \textbf{NNBTILQ027SBDIx} & 0.52 & pr              & 0.48 \\
2  & \textbf{capr}             & 0.68 & pr              & 0.58 & \textbf{capr}            & 0.51 \\
3  & \textbf{NNBTILQ027SBDIx}  & 0.75 & \textbf{capr}            & 0.58 & \textbf{NNBTILQ027SBDIx} & 0.56 \\
4  & HOUSTW           & 0.82 & AMDMUOx         & 0.80 & ISRATIOx        & 0.85 \\
5  & prfi\_gdp        & 0.85 & CES1021000001   & 0.81 & cli             & 0.85 \\
6  & AMDMUOx          & 0.87 & cli             & 0.87 & AMDMUOx         & 0.87 \\
7  & PERMIT           & 0.87 & IPNMAT          & 0.92 & CES1021000001   & 0.88 \\
8  & S\_P\_div\_yield & 0.87 & HOUST           & 0.95 & IPNMAT          & 0.89 \\
9  & HOUST            & 0.88 & PERMIT          & 0.96 & prfi\_gdp       & 0.91 \\
10 & PERMITMW         & 0.88 & HOUSTW          & 0.97 & IPDCONGD        & 0.96\\
				\hline \hline 
			\end{tabular}
		}
		\subcaption*{\textit{Notes}: Out-of-sample MSFEs for the $h$-quarter (cumulative) GDP growth rate, relative to an AR model. All models are estimated on recursive windows, using historical data vintages (see Section \ref{MAIN-subsec:realtime} of the paper for details). All MSFEs are computed over the period 1990Q1-2017Q4.}
		\label{tab:mse_ardl_var_realtime} 
	\end{table}

	\end{appendices}

	\end{document}